\RequirePackage{ifpdf}
\documentclass[letterpaper]{article}
\usepackage{jheppub}
\usepackage{xcolor}
\usepackage{amsmath}
\usepackage{epsfig}
\usepackage{caption}
\usepackage{subcaption}
\usepackage{soul}
\pdfoutput=1

\newcommand{\roughly}[1]{\mathrel{\raise.3ex\hbox{$#1$\kern-0.85em
\lower1ex\hbox{$\sim$}}}}
\newcommand{\lsim}{\roughly<}
\newcommand{\gsim}{\roughly>}

\def\nn{\nonumber}

\newcommand{\be}{\begin{equation}}
\newcommand{\bee}{\begin{equation}}
\newcommand{\ee}{\end{equation}}
\newcommand{\beea}{\begin{eqnarray}}
\newcommand{\eea}{\end{eqnarray}}
\newcommand{\bea}{\begin{eqnarray}}

\def\nott#1{\setbox0=\hbox{$#1$}                
   \dimen0=\wd0                                 
   \setbox1=\hbox{/} \dimen1=\wd1               
   \ifdim\dimen0>\dimen1                        
      \rlap{\hbox to \dimen0{\hfil/\hfil}}      
      #1                                        
   \else                                        
      \rlap{\hbox to \dimen1{\hfil$#1$\hfil}}   
      /                                         
   \fi}                                         %

\def\uxsl{\hbox{/\kern-.4000em$u$}}
\def\uxslsm{\hbox{\smaller/\kern-.5600em$u$}}
\def\pxpsl{\hbox{/\kern-.5000em$p$}}
\def\epssl{\hbox{/\kern-.5600em$\epsilon$}}
\def\delsl{\hbox{/\kern-.7000em$\nabla$}}
\def\lxpsl{\hbox{/\kern-.5600em$l$}}
\def\kxpsl{\hbox{/\kern-.5600em$k$}}
\def\qxpsl{\hbox{/\kern-.3900em$q$}}

\def\pref#1{(\ref{#1})}
\def\exd{{\rm d}}
\def\ol#1{{\overline{#1}}}

\def\cA{{\cal A}}

\def\cF{{\cal F}}

\def\cH{{\cal H}}
\def\cJ{{\cal J}}

\def\cL{{\cal L}}
\def\cM{{\cal M}}

\def\cO{{\cal O}}

\def\cR{{\cal R}}
\def\cS{{\cal S}}
\def\cT{{\cal T}}
\def\cU{{\cal U}}
\def\cV{{\cal V}}

\def\bfr{{\bf r}}

\def\bfn{{\bf n}}

\def\mfa{{\mathfrak a}}
\def\mfb{{\mathfrak b}}
\def\mfc{{\mathfrak c}}

\def\mfg{{\mathfrak g}}

\def\mfs{{\mathfrak s}}

\def\mfz{{\mathfrak z}}

\def\mfJ{{\mathfrak{J}}}

\def\ssA{{\scriptscriptstyle A}}
\def\ssB{{\scriptscriptstyle B}}
\def\ssC{{\scriptscriptstyle C}}
\def\ssD{{\scriptscriptstyle D}}
\def\ssE{{\scriptscriptstyle E}}
\def\ssF{{\scriptscriptstyle F}}

\def\ssH{{\scriptscriptstyle H}}

\def\ssM{{\scriptscriptstyle M}}
\def\ssN{{\scriptscriptstyle N}}
\def\ssP{{\scriptscriptstyle P}}

\def\ssT{{\scriptscriptstyle T}}
\def\ssU{{\scriptscriptstyle U}}

\def\ssX{{\scriptscriptstyle X}}

\def\PPN{{\scriptscriptstyle PPN}}
\def\BD{{\scriptscriptstyle BD}}

\def\SM{{\scriptscriptstyle SM}}

\def\QCD{{\scriptscriptstyle QCD}}

\def\TEV{{\scriptscriptstyle TEV}}

\title{Light Axiodilatons: Matter Couplings, Weak-Scale Completions and Long-Distance Tests of Gravity}

\author[1,2]{Philippe Brax,}
\author[2,3,4,5]{C.P.~Burgess,}
\author[6]{and F.~Quevedo}  

\affiliation[1]{Institut de Physique The\'orique, Universit\'e Paris-Saclay,
CEA, CNRS, F-91191 Gif-sur-Yvette Cedex, France.
}

\affiliation[2]{CERN, Theoretical Physics Department, Gen\`eve 23, Switzerland.}

\affiliation[3]{Department of Physics \& Astronomy, McMaster University, 1280 Main Street West, Hamilton ON, Canada.
}
\affiliation[4]{Perimeter Institute for Theoretical Physics, 31 Caroline Street North, Waterloo ON, Canada.
}

\affiliation[5]{School of Theoretical Physics, Dublin Institute for Advanced Studies,
 10 Burlington Road, Dublin, 
Ireland}

\affiliation[6]{DAMTP, University of Cambridge, Wilberforce Road,  Cambridge, CB3 0WA, UK.}

\date{\today}

\abstract{We consider the physical implications of very light axiodilatons motivated by a novel mechanism to substantially reduce the vacuum energy proposed in {\tt arXiv:2110.10352}. We address the two main problems concerning the light axiodilaton that appears in the low-energy limit, namely that the axion has a very low decay constant $f_a \sim $ eV (as read from its kinetic term) and that the dilaton is subject to bounds that are relevant to tests of GR once $\rho_{\rm vac} \lsim 10^{-80} M_p^4$. We show that eV scale axion decay constants need not be a problem by showing how supersymmetric extra dimensions provide a sample unitarization for axion physics above eV scales for which non-anomalous matter/axiodilaton couplings can really have gravitational strength, showing how naive EFT reasoning can mistakenly overestimates axion interactions at eV. When axions really do couple strongly at eV scales we identify the dimensionless interaction in the UV completion that is also O(1), and how axion energy-loss bounds map onto known extra-dimensional constraints. We find a broad new class of exact exterior solutions to the vacuum axiodilaton equations and knowledge of axiodilaton-matter couplings also allows us to numerically search for interior solutions that match to known exterior solutions that can evade solar-system tests. We find no examples that do so, but also identify potential new candidate mechanisms for reducing the effective dilaton-matter coupling to gravitating objects without also undermining the underlying suppression of $\rho_{\rm vac}$. }

\begin{document}

\maketitle

\section{Introduction}

When spinless particles are found to be very light they are usually Goldstone bosons for broken symmetries, at least approximately. Broadly speaking, there are two classes of Goldstone boson that arise in this way, depending on whether or not the underlying symmetry is compact or noncompact. 

For compact symmetries the dimensionless Goldstone field, $\mfa$, behaves as an angular variable that parameterizes a periodic direction in the scalar target space whose period (conventionally) is $2\pi$, and the approximate symmetry corresponds to shifts $\mfa \to \mfa + c$ for constant $c$ (or their nonabelian generalizations) as happens for axions. Scale transformations $\tau \to \lambda \tau$ for constant $\lambda$ (which can of course be regarded as shifts for the field $\ln\tau$) are simple examples of noncompact symmetries, for which the Goldstone direction is not periodic. 

Supersymmetry often bundles these two types of scalars together into complex combinations,\footnote{We here follow supergravity practice and call the real part of any such a field a `dilaton' and call its imaginary part an `axion'. } $\cT= \frac12(\tau + i \mfa)$, with the scaling and compact symmetries assembled into a larger $SL(2,R)$ group 
\be
   i\cT \to \frac{ic_1 \cT + c_2}{i c_3 \cT + c_4} \quad \hbox{with} \quad c_1 c_4 - c_2 c_3 = 1 \,.
\ee
Such scalars are rife in low-energy string vacua because these turn out to be riddled with scale, shift and duality symmetries on very general grounds \cite{UVShadows}. Mathematically such a complex scalar parameterizes the coset space $SL(2,R)/U(1)$ and the symmetry implies their kinetic term takes the form
\be \label{SL2Rkin}
  \cL_{\rm kin} = - \sqrt{-g} \; \mfz M_p^2 \frac{\partial \ol \cT \, \partial \cT}{(\cT+\ol \cT)^2} = - \sqrt{-g} \;\mfz M_p^2 \frac{(\partial \tau)^2 + (\partial \mfa)^2}{4\tau^2} \,,
\ee
where the Planck mass is related to Newton's constant by $M_p^{-2} = 8\pi G_\ssN$ and the order-unity constant turns out to be $\mfz=3$ in the cases of interest encountered below.

Normally the study of these scalars is regarded as being a purely theoretical exercise, relevant only at very high energies where supersymmetry might play a role. However recent developments \cite{YogaDE, Homeopathy, LowESugra} -- raise the possibility that they might survive to low energies if supersymmetry survives less broken in the gravitational sector than in the particle-physics sector (as indeed might be expected given that gravity naturally couples more weakly to any supersymmetry breaking sector). If so, axio-dilatons could be light enough to have practical implications for astrophysics and cosmology, whose potential possibilities and problems both hinge on the target-space interactions implied by \pref{SL2Rkin}. 

A perceived drawback of these models is their apparent need for UV completion at very low (eV) scales; a very practical obstruction to assessing their viability (such as whether they can survive the many constraints -- {\it e.g.}~energy-loss bounds -- that arise at higher energies). A purpose of this paper is to identify a natural class of UV completions, showing in particular how they can be used to explore high-energy constraints. This also allows us to clarify how the axiodilaton couples to ordinary matter, and what these interactions imply for tests of GR and other constraints. Along the way we find a broad new class of solutions to the axio-dilaton field equations. 

Before summarizing these results we first briefly recap the relevant features of \cite{YogaDE, Homeopathy} that we think motivate their study and define the problems that needs resolving.

\subsection*{The Yoga scenario}

Consider first the Yoga scenario of reference \cite{YogaDE}. In these models supersymmetry survives below the weak scale, but only in the gravity sector. The idea is to exploit the way this restricts how gravity responds to particle energies. The model borrows a universal feature common to the known string compactifications: an accidental and approximate classical scaling symmetry for which a complex field like $\cT$ contains the dilaton $\tau$. Corrections to scaling occur because the lagrangian arises as a series in powers of $1/\tau$, and the core idea behind the model  exploits a general mechanism\footnote{For aficionados: the structure of the potential arises because accidental scale invariance gives the leading terms a no-scale form \cite{NoScale}, whose breaking leaves the potential unusually shallow because of the `extended no-scale structure' mechanism described in \cite{UVShadows} (and first discovered within string compactifications in \cite{ExtendedNoScale}).} that ensures that these corrections first contribute to the scalar potential at one order higher in $1/\tau$ than one would naively expect. The scenario explores how large $\tau$ can be and whether this suppression can be signficant enough to be useful for the cosmological constant problem \cite{Weinberg:1988cp, Burgess:2013ara}. 

Although present for other reasons, the accidental scale invariance also makes $\tau$ couple to Standard Model fields only through a Brans-Dicke type \cite{Jordan, BransDicke, Dicke:1964pna, Brans} rescaling of the metric $\cL_\SM = \cL_\SM(\tilde g_{\mu\nu},\psi)$, where $\psi$ denotes a generic Standard Model field and $\tilde g_{\mu\nu} = A^2(\tau) \, g_{\mu\nu}$ with $A \propto \tau^{-1/2}$. Because of this the Higgs vev is proportional to $\tau^{-1/2}$ in the Einstein frame (for which $M_p$ is $\tau$-independent), and so the same is also true for all ordinary particle masses:\footnote{The QCD scale $\Lambda_\QCD \simeq M e^{-c/\alpha_s(M)}$ also scales as $\tau^{-1/2}$ when the reference UV mass scales like $M\propto \tau^{-1/2}$, ensuring {\it all} masses for ordinary (non-neutrino SM) particles scale in the same way.} $m_i/M_p \propto \tau^{-1/2}$. Neutrino masses can (but need not) be an exception: if they depend quadratically\footnote{This is true if {\it e.g.}~neutrino masses arise from the unique dimension-five SMEFT interaction of \cite{WeinbergDim5}, directly suppressed by $1/M_p$ rather than another $\tau$-dependent mass.} on the Higgs vev they instead satisfy $m_{\nu_a}/M_p \propto \tau^{-1}$. These scalings are broadly consistent with $m_i$ being at TeV scales and $m_{\nu_a}$ at sub-eV scales  (up to small dimensionless Yukawa couplings) if $M_p$ is the fundamental reference scale and $\tau \sim 10^{28}$. The electroweak and neutrino-mass hierarchies are also set by whatever physics fixes the size of $\tau$.

Like everything else, the scalar potential for $\tau$ arises as a series\footnote{Quantum breaking of scale invariance also allows the coefficients $V_i$ to depend logarithmically on $\tau$.} in $1/\tau$. In Planck units it has the form
\be \label{VtauSeries}
   V(\tau) = \frac{V_2}{\tau^2} + \frac{V_3}{\tau^3} + \frac{V_4}{\tau^4} + \cdots \,,
\ee
where supersymmetry of the gravity sector implies \cite{Komargodski:2009rz, Bergshoeff:2015tra, Dudas:2015eha, DallAgata:2015zxp, Schillo:2015ssx} $V_2$ arises as the perfect square 
\be \label{V2Wx}
   V_2 \propto |w_\ssX|^2 \,,
\ee
for some quantity $w_\ssX$, and so is strictly non-negative. This first term is also not particularly small, since its dependence on $\tau$ is consistent with the size that would be expected for a generic vacuum-energy contribution, $m^4$, given that $m \propto \tau^{-1/2}$. 
 
So far nothing particularly remarkable has happened. But because the dominant term (for large $\tau$) is a square, it likes to be minimized at zero if it should depend on another `relaxation' field,\footnote{$V_2$ vanishing dynamically resembles how charged scalars often seek the zeros of D-terms in supersymmetric models.} $\chi$, for which $w_\ssX(\chi) = 0$ has a solution for $\chi = \chi_0$. In practice $V_2$ typically is not minimized precisely at zero because the presence of higher powers of $1/\tau$ tend to drag $\chi$ away from the zero of $V_2$ ({\it e.g.}~$V_3$ turns out to be linear in $w_\ssX$). Instead $\chi$ is minimized where $w_\ssX \propto 1/\tau$, and so the minimum occurs where $V_2 \sim 1/\tau^2$ and $V_3 \sim 1/\tau$, making the potential at the minimum order $V_{\rm min} \sim 1/\tau^4$ once $\chi$ is minimized. The `natural relaxation' as $\chi$ seeks its minimum gives these `Yoga' models their name. 

Remarkably the resulting dependence on $\tau$ is consistent with $V_{\rm min} \sim v^4$ with energy scale $v \sim m^2/M_p$ where $m \sim M_p/\sqrt\tau$ is a typical Standard Model scale; the seed of an explanation for the famous numerology that the observed Dark Energy density is $v \sim M_\TEV^2/M_p$. Although relating all three of the electroweak, neutrino and Dark Energy hierarchies to the size of one field $\tau$ is tantalizing, why should $\tau$ be so large at its minimum? It is here that the dependence of $V_i$ on $\ln\tau$ plays a role, with \cite{YogaDE} showing how reasonable choices for this dependence can easily produce minima with $\tau_{\rm min} \sim 10^{28}$ given hierarchies amongst the parameters in $V_i$ that are only order $\ln\tau_{\rm min} \sim 65$.

It is of course the supersymmetry breaking masses $\Lambda_s \gsim 10$ TeV of any heavy superpartners of SM particles that are the most dangerous from the vacuum-energy point of view, and the main tension driving these models is to arrange parameters so that these superpartners can be heavy enough to have escaped detection while keeping a lid on the size of $V_{\rm min}$. In the best examples found in \cite{YogaDE} the vacuum energy turns out to be 
\be
   V_{\rm min} \sim \frac{\Lambda_s^4}{\tau_{\rm min} (\ln\tau_{\rm min})^5} \,,
\ee
and so $V_{\rm min} \sim 10^{-93} M_p^4$ if $\Lambda_s \sim 10$ TeV and $\tau_{\rm min} \sim 10^{28}$. In this expression the suppression by $1/\tau_{\rm min}$ is a consequence of the `extended no-scale structure' mechanism \cite{UVShadows, ExtendedNoScale} mentioned above, and the powers of $\ln\tau_{\rm min}$ are an accidental consequence of the stabilization mechanism used for $\tau$ \cite{YogaDE}. Although impressively small, this is not (yet) as small as the observed value: $10^{-120} M_p^4$. 

Improving on this is not our focus here however; we instead address the two main phenomenological problems implied by the Yoga framework, with a goal of identifying what the main remaining obstacles are (and hopefully suggesting model-building directions to overcoming them). We consider in turn the two issues that seemed in \cite{YogaDE} the most pressing: small axion decay constant and the dilaton's implications for tests of General Relativity (GR). Our conclusions are again mixed.

\subsection*{The axion problem}

The target-space interactions seen in \pref{SL2Rkin} appear to contain a problem when $\tau_{\rm min}$ is as large as required for the other hierarchies. Because the axion kinetic term is $M_p^2 \, (\partial \mfa)^2/\tau^2$ it predicts at face value the present-day axion decay constant to be $f_a \sim M_p /\tau_{\rm min}$. This makes $f_a$ similar in size to $v$ and $m_\nu$ (of order the eV scale) for the value $\tau_{\rm min} \sim 10^{28}$. 

This could be a problem if this is the scale the controls axion couplings to ordinary matter, and this is indeed what might be naively expected because once canonically normalized ($a := f_a \mfa$) the lagrangian 
\be\label{axion-matter0x}
  \cL_{\rm ax} =  - \frac{f_a^2}{2} \, \partial_\mu a \, \partial^\mu a -  \partial_\mu a \, J^\mu + \cdots
\ee
would predict non-renormalizable derivative axion-matter couplings of size $f_a^{-1} \partial_\mu a\,  J^\mu$. Such terms only make sense in an EFT that computes observables at energy $E$ in powers of $E/f_a$. Although it is tempting to conclude that the axion must be `strongly coupled' at these scales, what really happens is the derivative expansion breaks down for $E \gsim f_a$ and so the real size of the couplings must be understood in the UV completion that intervenes at eV energies. Such a completion is also required to see whether the model is consistent with any constraints on axion physics ({\it e.g.}~energy-loss bounds) that involve energies above eV scales. 
 
The point of view taken in \cite{YogaDE} is to agree that a UV completion must intervene at energies of order $f_a$, but to put aside the question of what this is in order to sort out the other naturalness and phenomenological issues that {\it can} be addressed at energies below eV scales. This is already nontrivial because Yoga model cancellations can be traced parametrically as functions of $\tau$, despite a full treatment of UV sensitivity requiring access to weak-scale energies. The viability of cancelling both $\tau^{-2}$ and $\tau^{-3}$ terms within $V$ can be addressed purely at low energies, as can other phenomenological issues like solar system tests and cosmology. 

We revisit the issue of UV completion in \S\ref{sec:UVcompletion} below, using extra dimensional models to pin down axion couplings more precisely\footnote{See also \cite{Companion}, that more broadly addresses to how UV naturalness issues -- like the quality problem -- are manifested for axions in extra-dimensional models that are {\it not} tied to a large extra-dimensional scenario.}. We consider three types of axions in such models, and we argue that one type of them can couple to ordinary matter with only gravitational strength (as do most KK modes) even though the kinetic term has the form \pref{SL2Rkin} with $\tau$ as large as $10^{28}$. The error in the naive estimate based on the lagrangian of \pref{axion-matter0} or \pref{axion-matter00} is to assume that the $\partial_\mu \mfa J^\mu$ term of \pref{axion-matter00} is independent of $\tau$, which the UV completion shows need not be true (at least for non-gauge interactions).\footnote{QCD-style axion-gauge couplings to dark gauge fields {\it are} possible (but not required) in this case, and would arise with strength $1/f_a$.} 

\subsection*{The Brans-Dicke problem}

The other problem starts with the observation that small $V_{\rm min}$ implies the dilaton $\tau$ is light:\footnote{Indeed this size is generic for {\it any} gravitationally coupled scalar that acquires its mass from $V$, so light scalars could be common if the potential's minimum is small.} $m^2_\tau \sim V_{\rm min}/M_p^2$. If $V_{\rm min}$ can be made small enough to describe the Dark Energy, the $\tau$ mass is of order the present-day Hubble scale. Once $V_{\rm min} \lsim 10^{-80} M_p^4$ we have $m_\tau^{-1} \gsim 10$ km, making the dilaton relevant to tests of GR and searches for new long-range forces. Furthermore, because particle masses are proportional to $\tau^{-1/2}$ the dilaton necessarily couples to matter as does a Brans-Dicke-like scalar, and does so with gravitational strength. Surely it must already be ruled out by such tests?

Possibly. As described in \cite{Homeopathy}, the derivative interactions between $\tau$ and $\mfa$ contained in \pref{SL2Rkin} complicate the predictions for tests of GR, provided that the axion $\mfa$ is also light enough to mediate a long-range force and also couples weakly to matter. To the extent that these couplings source the axion field and reduce the dilaton field they can help evade detection because observations are sensitive to the Weyl factor $A$ appearing in $\tilde g_{\mu\nu} = A^2 g_{\mu\nu}$ and this does not depend on the axion. To good approximation the axion field drops out of test-particle motion and so its presence tends to be missed by observational constraints. To decide whether this escape mechanism is actually employed by real systems requires matching exterior solutions to the solutions within the source's interior (something that we address in this article), and this in turn requires more information about how the axiodilaton couples to matter.

\subsection*{So what's new?}

This brings us to the purpose of the paper you are now reading. Lack of information about axiodilaton-matter couplings at ordinary energies obstructs progress on both of the above problems. We therefore first identify UV completions that can be valid up to electroweak scales, doing so by pursuing the suggestion made in \cite{YogaDE} that large extra dimensions could intervene at eV scales. If so, only two dimensions can be this large\footnote{More dimensions than the minimal two can arise (as usual) provided they only do so above TeV energies.} without already having been detected\footnote{Recently, motivated by some swampland conjectures, an interesting dark energy proposal was put forward \cite{mvv}, for which the case is made for a single large extra dimension rather than two. It is not clear how, in this scenario, loops of Standard Model particles are cancelled to keep the cosmological constant small, but it may be interesting to explore any potential connection with our proposal.}, and even these can only be this large without escaping detection if all Standard Model particles are trapped on a 4D space-filling brane situated somewhere within the two large dimensions \cite{LED, LED2, LED3}. 

Indeed the entire framework wherein nonsupersymmetric Standard Model physics couples to a supersymmetric gravity sector is naturally captured in this framework if the Standard Model brane badly breaks supersymmetry but the extra-dimensional bulk is otherwise supersymmetric. In this case supersymmetry in the gravity sector is only broken by boundary conditions and the lightest gravitino is expected to have a mass of order the Kaluza Klein (eV) scale (in agreement with the gravitino mass $m_{3/2}$ that was also found in a purely 4D way within the Yoga setup). Extra dimensions also provide a microscopic interpretation for the large value for $\tau$: it encodes the large volume of the extra dimensions:\footnote{Ref.~\cite{YogaDE} provided arguments why the simplest connection between $\tau$ and $\cV_2$ could have problems, and \S\ref{ssec:AsymmetricYC} below addresses why these need not be fatal.} $\cV_2 \sim (ML)^2 \sim 10^{28}$ where $L$ is the extra-dimensional radius and $M$ the 6D gravity scale. 

In this way of thinking the natural UV extension of Yoga models is supersymmetric large extra dimensions (SLED), which was indeed initially proposed \cite{SLED} with Dark Energy density in mind. Much is known about UV sensitivity in this framework and can be carried over in whole cloth to Yoga models. For instance UV scales do not gravitate as in 4D, with UV on-brane contributions typically curving the transverse extra dimensions rather than the four on-brane dimensions visible to brane-bound cosmologists \cite{Carroll:2003db, SLED, UVSLEDb}. Supersymmetry of the gravity sector (the bulk) similarly suppresses UV contributions from other branes or from elsewhere in the extra-dimensional bulk \cite{UVSLEDBulk}. The requirement that Standard Model particles live on a 4D space-filling brane also means that most non-gravitational predictions go through as in four dimensions, and so are as captured by the Yoga-model lagrangian even at energies above eV scales. It is only for high-energy {\it gravitational} processes that the full extra-dimensional machinery is really required. 

We find several interesting consequences of this UV picture. As remarked elsewhere \cite{Axiverse} axions are generic in supersymmetric extra-dimensional models, for example arising as KK modes for the 2-form gauge fields that commonly appear as part of the extra-dimensional graviton supermultiplet. \S\ref{sec:UVcompletion} revisits how extra dimensions unitarize axion interactions, sometimes leaving them with gravitational-strength interactions with ordinary matter despite the apparent breakdown of the $E/f_a$ expansion at eV scales. 

Our main line of inquiry asks what kinds of UV matter couplings can ameliorate the Brans-Dicke and axion problems described above. In particular, we test whether the interior solutions suggested by the simplest axion-matter couplings actually match to the exterior solutions found in \cite{Homeopathy} that evade solar system tests of gravity. We identify conditions that the axion-matter couplings must satisfy in order for this to happen, and report on extensive numerical searches for successful solutions using the simplest types of axion-matter couplings, none of which is ultimately successful but which also identify new ways in which axion-matter couplings can act to suppress the effective dilaton charges of macroscopic objects. 

In one promising approach it is a chameleon-type mechanism for the axion that gets transferred by axio-dilaton self-interactions to reduce the apparent dilaton charge. But we do not yet have a mechanism that would evade all bounds and so we still consider it an open challenge to find a way for the Yoga relaxation mechanism to evade solar-system constraints. Our search suggests several further directions for how to find phenomenologically successful couplings, and regard this type of model building to be a constructive reformulation of the cosmological constant problem since it leads to directions that have not yet been fully explored. Given the magnitude of the original problem (dark energy), recasting research along these lines seems worthwhile.

The rest of the paper is organized as follows. \S\ref{sec:UVcompletion} shows how the axiodilaton lagrangian arises by explicit dimensional reduction of extra-dimensional models and how such models provide a natural framework for exploring their properties at scales $E \gsim f_a$. It is also argued how the large values required for $\tau$ can be plausibly achieved within this framework if two extra dimensions are larger than the others and close to their upper allowed size since in this case the two large dimensions have volume (in fundamental units) $\cV_2\sim 10^{28}$.

\S\ref{ssec:Unitarization} then identifies three ways ($S$-type, $T$-type and $\Phi$-type) axions can arise from the particularly rich case of extra-dimensional two-form fields, $B_{\ssM\ssN}$, that are generic to supersymmetric models. Their possible couplings to on-brane (Standard Model) degrees of freedom are computed by direct dimensional reduction, and this shows why the $T$-type axion can have low-energy matter couplings proportional to $\cF^{-1}$ with effective decay constant\footnote{Having UV physics intervene with only weak couplings is not uncommon for EFTs \cite{EFTBook}, and underlines why it can be perverse to call the breakdown of the low-energy expansion `strong coupling' (as is sometimes done in the literature).} $\cF \sim M_p$ rather than $f_a$. The naive strong-coupling argument given above based on \pref{axion-matter0x} is mistaken in this case because it is naive about how factors of the large volume $\cV_2$ appear in the interaction terms at low energies. 

Higher-dimensional gauge symmetries turn out to preclude these Planck-suppressed $T$-axion/matter couplings from including QCD-like axion-gauge interactions. The other two types of axions can couple to QCD, and $\Phi$-type axions in principle can do so with an effective decay constant that can be large enough to play a role in the strong CP problem. The natural size for $S$-type axion/matter couplings really is $1/f_a$, as it happens, but such couplings need not be present. It is not implausible for such models to have dark (bulk) gauge sectors and axions can couple to these with strength $1/f_a$. These match to $\cO(1)$ dimensionless effective couplings within the higher-dimensional UV completion. 

The remaining sections then pivot to using this framework to explore several phenomenological issues, with \S\ref{sec:UVpheno} focussing on higher energy issues that require the UV completion. This section argues that inclusive axion energy-loss bounds generically coincide with the standard extra-dimensional energy loss constraints that require the scale of 6D gravity to be above tens of TeV, and discusses which kinds of predictions depend on extra-dimensional details and which do not. The 4D Yoga-model perspective turns out also to help understand why SLED models were promising but ultimately not completely successful in accounting for the Dark Energy density, and why Yoga models might do better. 

\S\ref{sec:IRscreening}  turns to the core phenomenological problems these models face: constraints coming from solar-system tests of gravity in scenarios where $V_{\rm min}$ is low enough that the dilaton becomes light enough to mediate a macroscopically long-range force. It does so under the assumption that the vacuum does not break the axion shift symmetry (though the matter-axion couplings within a gravitating source might). We find a broad new class of exact solutions to the classical axiodilaton field equations external to a source, including those that are not rotationally invariant and so that can capture effects like multipole moments. We use these solutions to show why shift symmetry breaking by matter-axion couplings is a necessary condition for modifying the effective dilaton-matter couplings along the lines proposed in \cite{Homeopathy}, and match these to a broad class of numerically generated interior solutions to see if effective matter-dilaton couplings can evade detection. We do not identify any that succeed in doing so. Our results are summarized in \S\ref{sec:Conclusions} and an appendix contains an analytic limit for the interior solutions that complements our numerical searches in the main text. 

\section{UV completions}
\label{sec:UVcompletion}

This section explores the types of axion couplings that are inherent in UV completions that involve supersymmetric large extra dimensions, and why they resolve the problems associated with decay constants in the eV regime. We identify two main types of axion structure that emerge ($T$-type and $S$-type axions) and show why both lead to physical modes that in reality couple only with gravitational strength \footnote{In general string compactifications there are several possibilities worth exploring in more detail, depending if the axion comes from a complex structure or K\"ahler modulus, from 2,3,4-forms and in IIB from the original 10D axion. It also depends on having the Standard Model on  branes of different dimensionalities. See for instance \cite{michele,joe}. }. The section closes with a discussion of why extra-dimensional models can avoid the constraints discussed in \cite{YogaDE} that naively seemed to preclude there being extra-dimensional UV completions.  

\subsection{Axion unitarization}
\label{ssec:Unitarization}

Axions whose kinetic terms have the form \pref{SL2Rkin} with $\tau \simeq \tau_{\rm min} \sim 10^{28}$ seem to imply a low-energy axion/matter lagrangian of the form
\be\label{axion-matter0}
  \cL_{\rm ax} =  - \frac{f_a^2}{2} \, \partial_\mu \mfa \, \partial^\mu \mfa -  \partial_\mu \mfa \, J^\mu + \cdots
\ee
with $J^\mu$ being a collection of lowest-dimension currents built from SM particles and decay constant $f_a = M_p/\tau_{\rm min} \sim 1$ eV. If required to be gauge invariant the currents $J^\mu$ first arise with mass dimension 3 involving either fermions or the Higgs doublet:
\be \label{Jpieces}
  J^\mu_\ssF =   \ol\psi \gamma^\mu g_{\ssF} \psi  \quad \hbox{and} \quad J^\mu_\ssH = ig_\ssH \Bigl[(D^\mu \cH^*) \cH - \cH^* (D^\mu \cH)\Bigr] \,,
\ee
for some Dirac/flavour matrix $g_{\ssF}$ and Higgs coupling $g_\ssH$. 

In the case of gauge bosons the lowest-dimension operators are not gauge invariant, but in some circumstances these can also be used to build gauge invariant interactions. The two lowest-dimension contributions of this type consist of the gauge potential $A_\mu$ and the Hodge dual of the Chern-Simons form $\omega_{\mu\nu\lambda}$. For abelian fields\footnote{For simplicitly we write explicitly only the abelian case, but the nonabelian generalization is straightforward.} these take the explicit form 
\be \label{JpiecesA}
    J_\ssA^\mu = g_\ssA \, A^\mu \quad\hbox{and} \quad J_\ssC^\mu = \frac{g_\ssC}{2} \epsilon^{\mu\nu\lambda\rho} A_\nu F_{\lambda\rho} \,.
\ee
Interactions like $\partial_\mu \mfa \, J_\ssA^\mu$ can come as parts of $(\partial_\mu \mfa + g_\ssA A_\mu)(\partial^\mu \mfa + g_\ssA A^\mu)$ and so arise when the axion shift symmetry is gauged. Their presence indicates gauge-field mass acquisition through the Stueckelberg mechanism. The Chern-Simons interaction is similarly seen to be gauge invariant by integrating by parts to rewrite $\partial_\mu \mfa\, J_\ssC^\mu$ as proportional to $\mfa \, \epsilon^{\mu\nu\lambda\rho} F_{\mu\nu} F_{\lambda\rho}$. Such interactions represent anomalies in global axion shift symmetries or contributions to anomaly cancellation if the axion shift symmetry is gauged. The Chern-Simons current $J_\ssC^\mu$ has straightforward generalization to nonabelian gauge fields.

Once expressed in terms of the canonically normalized field $a = f_a \mfa$ we have
\be\label{axion-matter00}
  \cL_{\rm ax} =  - \frac{1}{2} \, \partial_\mu a \, \partial^\mu a -  \frac{1}{\cF} \, \partial_\mu a J^\mu \,,
\ee 
with $\cF = f_a$. When $J^\mu$ has dimension (mass)${}^3$ this is a nonrenormalizable interaction which standard EFT reasoning argues should be interpreted as part of a low-energy derivative expansion. Because the dimensionless expansion is in $E/f_a$ this interpretation breaks down at energies $E \gsim f_a \sim 1$ eV, invalidating use of this low-energy EFT at higher energies. Something must intervene at or below $f_a$ to enable predictions at ordinary energies and we here argue that two large extra dimensions provide a simple and plausible example of what this could be. This section focusses specifically on how the above argument changes if the UV completion at these scales is extra-dimensional and shows why in this case a more reliable estimate for the size of axion-matter couplings reveals them to be Planck suppressed.

The point is most easily made using an explicit example, so consider the specific instance where the axion arises as a low-energy mode of an antisymmetric Kalb-Ramond field $B_{\ssM\ssN}$ within two extra dimensions. This system is known to produce the required $SL(2,R)$-invariant form for the axio-dilaton lagrangian used in \cite{YogaDE, Homeopathy}.

\subsubsection*{Axiodilatons from extra dimensions} 

Consider first how axions arise from higher-dimensional 2-form gauge potentials, $B_{\ssM\ssN}$. For concreteness' sake we do so assuming two large extra dimensions, though we also distinguish how axion properties differ if they arise from other smaller dimensions. For simplicity take the extra-dimensional metric to have the following product form:

\be \label{6Dmetric}
  \exd \tilde s^2 = \tilde g_{\ssM\ssN} \, \exd x^\ssM \exd x^\ssN = \frac{\cV_{2}^{(0)}}{\cV_2} \, g_{\mu\nu}(x) \,\exd x^\mu \exd x^\nu + \cV_2\, \hat g_{mn}(y) \, \exd y^m \exd y^n \,,
\ee
where the metric $\hat g_{mn}$ satisfies $\int \exd^2y \sqrt{\hat g} = M^{-2}$ so that $\cV_2 = (ML)^2$ is the extra-dimensional volume $L^2$ in units of a UV scale $M$. In practice $M$ is the extra-dimensional Planck scale, related to the 4D Planck scale by 
\be
  M_p^2 = M^4 L^2 = M^2 \cV_2^{(0)} \,,
\ee
where $\cV_{2}^{(0)}$ denotes the volume's present-day value, since $\cV_2$ (and $L$) are low-energy 4D fields that can vary in space and time. For two large extra dimensions $M \sim 10$ TeV and $\cV_2^{(0)} \sim 10^{28}$, which we will see is ultimately the origin of the large {\it vev} for the field $\tau$. The factor $\cV_2^{(0)}/\cV_2$ in \pref{6Dmetric} ensures the the metric $g_{\mu\nu}$ is the 4D Einstein-frame metric, and does so without changing overall units (because $\cV_2^{(0)}/\cV_2 = 1$ at present).  

There are three ways that an axion can arise from a 2-form field in this kind of setup:
\begin{enumerate}
\item It can be the zero-mode of the purely extra-dimensional components $B_{mn}(x,y) = \mfb(x) \, \omega_{mn}(y)$, where $\omega_{mn} \propto \hat\varepsilon_{mn}$ is proportional to the volume form for the two large extra dimensions built from the metric $\hat g_{mn}$. We call this the $T$-type universal axion.
\item It can be the zero-mode of purely extra-dimensional component, $B_{ab}(x,y,z) = \Phi(x,y) \,\omega_{ab}(z)$, within some smaller higher dimensions beyond the six written explicitly in \pref{6Dmetric} (such as could happen if the 6D theory were the low-energy limit of a 10- or 11-dimensional string vacuum). In this case $\omega_{ab}$ is a harmonic form within these smaller extra dimensions. Such an axion appears in the 6D theory directly as a 6D scalar whose zero-mode $\Phi(x,y) = \mfb(x)$ in the compactification to four dimensions is the axion of the 4D world, and whether such field arise in a given compactification to 6D is a model dependent issue. We call these the $\Phi$-type axions.
\item It can arise as the zero-mode of the purely four-dimensional part $B_{\mu\nu}(x,y) = b_{\mu\nu}(x)$, which in four dimensions dualizes to a scalar with $H_{\mu\nu\lambda} = \partial_\mu b_{\nu\lambda} + \cdots \propto \epsilon_{\mu\nu\lambda\rho} \partial^\rho \mfa$.  
We call this `dual' axion the `$S$-type' universal axion to distinguish it from the previous two cases.
\end{enumerate}
The harmonic form appearing in items 1 and 2 usually satisfies a quantization condition for which $\oint_\ssC \omega$ is a pure number when integrating over a 2-cycle $C$, and so $\omega_{ab} \propto \cV_\ssC^{-1}$ where $\cV_\ssC$ is the dimensionless volume of $C$. For $T$-type axions this means $\omega_{mn} = k \hat \varepsilon_{mn}$ with $k \propto \cV_2^{-1}$.

Two facts are central to fixing the size of axion-matter couplings. First, ordinary matter must be trapped on a 4-dimensional brane in order for the large extra dimensions to have escaped experimental observation \cite{LED, LED2, LED3}. Second, the kinetic energy of a higher-dimensional 2-form potential often involves more than these four brane dimensions; for concreteness we focus on the 2-form field that lives in the gravity supermultiplet and so which lives in the full extra-dimensional `bulk':
\be \label{BMNkin}
   S_{\rm kin} = -\frac{M_{(\ssD)}^{D-4}}{2\cdot 3!} \int \exd^\ssD x \; \sqrt{- \tilde g_{(D)}} \; e^{-2\mfs} \, H_{\ssM\ssN\ssP} H^{\ssM\ssN\ssP} \,,
\ee
where $M_{(\ssD)}$ is a UV scale (equal to $M$ in 6D) $\tilde g_{(D)}$ denotes the $D$-dimensional determinant of the higher-dimensional metric $\tilde g_{\ssM\ssN}$ within the extra-dimensional Einstein frame and $H_{\ssM\ssN\ssP} = \partial_\ssM B_{\ssN\ssP} + \hbox{(cyclic)}$ is the Kalb-Ramond field strength. Here $\mfs$ is the 6D dilaton that often also arises as part of the higher-dimensional gravity supermultiplet.

The relevance of these two facts is easiest to see for item (1) above, where $\mfb$ arises from $B_{mn}$ in the two large extra dimensions. To start with, dimensionally reducing the $D=6$ version of \pref{BMNkin} gives the $\mfb$ kinetic term, including the following dependence on $\mfs$ and $\cV_2$:
\be \label{TaxKinDR}
  M^4 \int \exd^2y \sqrt{-\tilde g_{(6)}} \;e^{-2\mfs} \tilde g^{\mu\nu} \tilde g^{mn} \tilde g^{pq} \partial_\mu B_{mp} \partial_\nu B_{nq} \propto M^2 \cV_2^{(0)} \sqrt{-g}\; e^{-2\mfs} \cV_2^{-2} g^{\mu\nu} \partial_\mu \mfb \,\partial_\nu \mfb \,.
\ee
This is consistent with the axion kinetic term $\sqrt{-g}\,M_p^2\, (\partial \mfb)^2 /\tau^2$ of \pref{SL2Rkin} with $\tau = \cV_2 \, e^\mfs$. 

\subsubsection*{Axion-matter couplings}

To couple $\mfb$ to matter in the 6D effective theory we require a generally covariant and gauge invariant interaction that couples $H_{\ssM\ssN\ssP}$ to matter localized on a space-filling 4D brane, containing in particular the components $H_{\mu mn} \ni \partial_\mu B_{mn}$. A term linear in $H$ that contains $\partial_\mu \mfb$ can be built in a covariant way using the 6D Hodge dual ${}^\star H$, pulled back to the brane and wedged with a matter current $J_\mu$:
\be \label{bwedgeint}
   S_{\rm int} = \int_{\Sigma_b} h_t(\mfs) \;  {}^\star H  \wedge J  \,,
\ee
where $h_t(\mfs)$ allows for a possible dependence on the dilaton. Notice $J$ cannot here be a Chern-Simons current as in \pref{JpiecesA} because $S_{\rm int}$ in this case is not gauge invariant.

Isolating the contribution involving $\partial_\mu \mfb$ and absorbing dimensionless numerical factors into the current $J_\mu$ leads to the following dependence on $\cV_2$ and $\mfs$:
\be
     \sqrt{-\tilde g_{(4)}} \; h_t(\mfs) \, \tilde g^{\mu\nu} \tilde \epsilon^{mn}(y_b) \, \partial_\mu B_{mn}(x,y_b) J_{\nu}(x) 
    = \cV_2^{(0)} \sqrt{-g} \; h_t(\mfs) \, \cV_2^{-2} g^{\mu\nu} \partial_\mu \mfb(x) J_{\nu}(x)   \,,
\ee
where $y_b$ is the extra-dimensional brane position and $\tilde\epsilon_{mn}$ is the extra-dimensional volume form built using the metric $\tilde g_{mn}$. Combining kinetic and interaction terms gives -- for the special case $h_t(\mfs) = e^{-2\mfs}$ -- the following terms in the 4D Einstein-frame effective action
\be \label{TtypeKin}
   S_{\rm eff} = \int \exd^4x \, \sqrt{-g} \;  \frac{M_p^2}{\tau^2}\left[ (\partial \mfb)^2 + \frac{\partial_\mu\mfb J^\mu}{M^2} \right]
\ee
where $\tau := \cV_2 \, e^\phi$ as before. 

Canonically normalizing by rescaling $b = M_p \, \mfb/\tau_{\rm min}$ -- with $\tau_{\rm min} = \langle \tau \rangle \propto \cV_2^{(0)}$ -- then produces a lagrangian of the form \pref{axion-matter00} but with
\be\label{Fdef}
   \cF \sim \frac{M^2 \tau_{\rm min}}{M_p} 
   \sim M_p \,,
\ee
rather than $f_a$. As is typical for bulk fields, each KK mode within $B_{mn}$ couples with gravitational strength. The error leading to the earlier conclusion $\cF = f_a$ lies in ignoring the $\cV_2$ dependence that the interaction also inherits from the higher-dimensional metric. 

For the $\Phi$-type fields, the situation is more model dependent since they may or may not be localized on the brane (depending on the brane dimensionality), couplings to Chern-Simons currents (as in \pref{JpiecesA}) can be gauge invariant and the coupling to matter may be stronger (because the volume $\cV_\ssC$ of the relevant cycle can be much smaller). Contrary to the $T$ type axion (whose absence of a Chern-Simons coupling requires it to be an ALP), $\Phi$-type axions may be QCD-like. 

The story for the $S$-field axion is interestingly different, with strong matter couplings just as the naive argument suggests, but in this case for the scalar theory that is dual to the one obtained by dimensional reduction. The volume-dependence of the dimensionally reduced kinetic term is
\be \label{hmunurhokin}
   S_{\rm kin} \ni
   -\frac{M^4}{2\cdot 3! M_p^2} \int \exd^4x \; \sqrt{-g} \; e^{-2 \mfs}\,\cV_2^2 \, h_{\mu\nu\lambda} h^{\mu\nu\lambda} \,,
\ee
where $h_{\mu\nu\lambda} = \partial_\mu b_{\nu\lambda}$ + (cyclic). To dualize we impose (in 4D Einstein frame) the Bianchi identity $\exd h = \Omega$ using a Lagrange-multiplier field $\mfa$; supplementing \pref{hmunurhokin} with
\be
   S_{\rm bi} = \frac{1}{3!} \int \exd^4 x  \sqrt{-g}\; \mfa \, \epsilon^{\mu\nu\lambda\rho} \Bigl(M^2 \, \partial_\mu h_{\nu\lambda\rho} - \Omega_{\mu\nu\lambda\rho} \Bigr) \,.
\ee
Here $\Omega$ is a gauge-invariant closed 4-form -- {\it i.e.}~one that satisfies $\exd \Omega = 0$ -- built from gauge fields and the metric -- that typically lives in the bulk as does $B_{\ssM\ssN}$. In practice we are mostly interested in the case $\Omega = F \wedge F$ for $F$ a gauge field strength, though for bulk fields this is a dark gauge sector. 

Lowest-dimension couplings to matter currents localized on a 4D brane have the form
\be \label{awedgeint}
   S_{\rm int} = \int_{\Sigma_b} h_s(\mfs) \;  H  \wedge J  \,,
\ee
where $h_s(\mfs)$ again allows for a possible dependence on the dilaton. In this case gauge invariance allows $J$ to be a Chern-Simons current as in \pref{JpiecesA} but only if the dimensionless coupling function $h_s(\mfs) = h_{s0}$ is $\mfs$-independent.

The functional integral of $S_{\rm kin} + S_{\rm bi}$ with respect to $h_{\mu\nu\lambda}$ and $\mfa$ is equivalent to integrating $S_{\rm kin}$ with respect to $b_{\mu\nu}$ because the integral over $\mfa$ imposes the Bianchi identity, and so allows the integral over $h_{\mu\nu\lambda}$ to be traded for one over $b_{\mu\nu}$. The dual result is obtained by instead performing these integrals in the opposite order; first performing the gaussian integral over $h_{\mu\nu\lambda}$ (see \cite{Companion} for details). 
The result for the dual lagrangian then becomes
\be \label{StypeDual}
   S_{\rm dual} = - \int \exd^4x \, \sqrt{-g} \; \left[ \frac{M_p^2}{2\cV_2^2}\,  e^{2 \mfs} \, D_\mu \mfa \, D^\mu \mfa +  \frac{1}{3!}  \, \mfa \, \epsilon^{\mu\nu\lambda\rho}  \Omega_{\mu\nu\lambda\rho} \right]\,,
\ee
where $D_\mu \mfa = \partial_\mu \mfa + h_s(\mfs) \, e^{-2\mfs} \cV_2^2 J_\mu/M_p^2$. When $h_s$ is a constant the kinetic term again has the form \pref{SL2Rkin}: $\sqrt{-g}\,(\partial\mfa)^2/\sigma^2$, but this time with $\sigma = \cV_2\, e^{-\mfs}$. 

We note in passing that this kinetic term combines with the kinetic term for $\mfb$ found in \pref{TtypeKin} and the kinetic terms for the fields $\tau = \cV_2 \, e^\mfs$ and $\sigma = \cV_2\, e^{-\mfs}$ (obtained from the 6D Einstein action and dilaton kinetic term) into the form \pref{SL2Rkin}; a form captured by the low-energy 4D supergravity K\"ahler potential 
\be \label{SSKPot}
   K/M_p^2 = - \ln (\cS+\ol \cS) - \ln (\cT + \ol \cT) = - \ln (\sigma \tau) = - 2 \ln \cV_2
\ee
where $\cS = \frac12(\sigma + i \mfa)$ and $\cT = \frac12(\tau+ i \mfb)$. These kinetic terms reveal that both types of axions have the form \pref{SL2Rkin}: with naive decay constants, $f_a = M_p/\sigma$ and $f_b = M_p/\tau$, and so both are eV in size when $\cV_2 \sim 10^{28}$, consistent with how the 4D EFT breaks down at the order eV Kaluza-Klein scale associated with large dimensions. 

An important difference between \pref{StypeDual} and our earlier examples is that interaction terms like $\mfa\, F \wedge F$ or \pref{awedgeint} do not involve the metric and so does not contain hidden factors of $\cV_2$. As a result the naive argument for the size of $\cF$ is in this case correct: comparing the kinetic and interaction terms reveals the physical coupling has the form of \pref{axion-matter00} with strength $\cF = f_s = M_p/\sigma$. This time the theory \pref{StypeDual} {\it does} have order-unity couplings at $E = f_s$. In the UV completion these match to dimensionless couplings like $h_{s0}$ appearing in interactions like \pref{awedgeint}. 
 
\subsection{Large $\tau$ from asymmetric compactifications}
\label{ssec:AsymmetricYC}

The possibility these models UV complete at eV energies to supersymmetric large dimensions was considered in \cite{YogaDE}, though on first inspection this seemed difficult to do, at least within the context of Type IIB supergravity. In this section we sketch why we no longer regard the perceived difficulties described in \cite{YogaDE} to be a problem.\footnote{Since our purpose here is only to identify mechanisms, we do not try to construct a fully modulus-stabilized theory (as would be required if we were to push the upper UV limit up past the weak scale into the fullly 10D string regime).}

The root of the problem described in \cite{YogaDE} was this: phenomenology prefers a 4D K\"ahler potential of the form $K = - 3 \ln (\tau + \cdots)$ where $\tau \sim 10^{28}$. We would like to build such a $\tau$ from the basic hierarchy of the (at most) two large dimensions that must arise up to TeV energies if extra dimensions are already to become relevant at eV energies.\footnote{Warping was also explored as a potential additional source of hierarchy in \cite{YogaDE}.} At first sight the numerology is promising because taking $1/L \sim 1$ eV and a fundamental extra-dimensional UV scale $M_s \sim 100$ TeV implies the dimensionless length scales are $M_s L = 10^{14}$ and so the dimensionless extra-dimensional 2D volume is $\cV_2 = (M_s L)^2 = 10^{28}$. Since dimensional reduction in 6D implies $M_p^2 = M_s^2 \cV_2$ the volume $\cV_2$ cannot be much larger than this without $M_s$ becoming too small (or $L$ becoming too large) to have been missed in experiments. 

The perceived difficulty arose once the precise connection is made between $\cV_2$ and $\tau$. In very many cases the low-energy K\"ahler potential found by dimensional reduction is given by an expression like \pref{SSKPot}, with 
\be \label{KSS}
   K = - 2 \ln \cV  
   \,,
\ee
but if this is identified with $K = -3 \ln\tau$ it implies $\tau = \cV^{2/3}$ (making $\tau\sim 10^{18}$ at most -- and so too small -- given $\cV \sim \cV_2 \lsim 10^{28}$). 

We now argue why some compactifications seem likely to evade this problem. To do so it is useful to consider the size of the moduli that are encountered when compactifying higher-dimensional theories on tori. For instance toroidal compactifications of 6D supergravity on a 2-torus give a K\"ahler potential of the form 
\be \label{STUK}
   K = - \ln(stu) \,,
\ee
(plus possibly other moduli) where $s \propto L_1 L_2 e^{-\mfs}$ and $t \propto L_1 L_2 e^\mfs$ can be as given above (where $L_i$ are the lengths of the torus' two fundamenal cycles), while $u$ is the torus' complex-structure modulus and so is proportional to $L_1/L_2$. \pref{STUK} agrees with \pref{SSKPot} in the simplest case where $L_1 = L_2 = L$ and so $u$ is order unity, but also suggests that we could get what we want if there were other fields (like $u$) that were of the same size as $s$ and $t$. 

To see where such fields might come from, for concreteness' sake consider extending into the far UV to include two more extra dimensions (as part of a fuller 10 dimensional theory\footnote{See for instance the discussion on the expressions for the string dilaton, K\"ahler and complex structure moduli in toroidal compactifications of the different string theories in \cite{luisbook}.}), compactified on the product of two 2-tori,\footnote{Tori should just be regarded as illustrative here, whose purpose is simply to show concretely how other large moduli might arise given only two large dimensions. In practice extensions further into the UV are likely to involve compactifications on other geometries, possibly with similar K\"ahler potentials, but with more explicit modulus stabilization (see {\it e.g.}~\cite{STUCY}).} In many situations such compactifications generate K\"ahler potentials for the geometrical moduli of the form
\be
  K = - \ln (\cS+\ol \cS) - \sum_{i=1}^2 \ln (\cT_i + \ol{\cT}_i) - \sum_{i=1}^2 \ln(\cU_i + \ol{\cU}_i) \,.
\ee
The fields $\cS$, $\cT_i$ and $\cU_i$ have different expressions in terms of the underlying length scales in different kinds of geometries, but we assume the $t_i = \cT_i + \ol \cT_i$ depend on the volume moduli for each torus (as found above for $t$ and the volume of the single torus in 6D) and that the $u_i = \cU_i + \ol \cU_i$ are the complex-structure moduli of the two 2-tori. If we denote the toroidal radii for the two tori by $(L_1,L_2)$ and $(L_3,L_4)$, then their volume moduli scale with lengths as $t_1 \propto L_1 L_2$ and $t_2 \propto L_3 L_4$ while the complex-structure moduli are $u_1 \propto L_1/L_2$ and $u_2 \propto L_3/L_4$.

We now ask how big $K$ can be if only two of the dimensions have a large length $L$ and the rest have the much smaller length $\ell$. Consider first the simplest case where both sides of one of the 2-tori is much bigger than both sides of the other 2-torus: $L_1 = L_2 = L$ and $L_3 = L_4 = \ell$. In this case $t_1 \sim L^2$ and $t_2 \sim \ell^2$ while the $u_i$ are both order unity. Assuming $s \propto L^2$ as before the argument of the logarithm in $K$ has size
\be
    s \left( \prod_{i=1}^2 t_i \right) \left( \prod_{i=1}^2 u_i \right) \propto L^2 (L^2 \ell^2) (1) = L^4 \ell^2 \qquad \hbox{(Case I)} \,,
\ee
and so $K = - \ln(s t_1 t_2 u_1 u_2) \sim - 3 \ln(L^{4/3})$. This is a specific instance of the generic situation discussed above, where $\tau \sim \cV_2^{2/3} \sim L^{4/3}$.

But if we instead assume $L_1 = L_3 = L$ and $L_2 = L_4 = \ell$ then we have $t_1 \sim t_2 \sim L \ell$ and $u_1 \sim u_2 \sim L/\ell$. In this case we instead have 
\be
    s\left( \prod_{i=1}^2 t_i \right) \left( \prod_{i=1}^2 u_i \right) \propto L^2(L \ell)^2 \left( \frac{L}{\ell} \right)^2 = L^6  \qquad \hbox{(Case II)}\,,
\ee
for which $K =  - \ln(s t_1 t_2 u_1 u_2) \sim - 3 \ln(L^{2})$ and so $\tau \sim \cV_2 \sim L^2 \sim 10^{28}$ can be possible.

We can be  explicit in toroidal orientifold models of type IIA and IIB string compactifications as discussed in chapter 12 of \cite{luisbook}. Denoting the radii of each of the three 2-tori by $R^i_\alpha $ with $i=1,2,3$ labelling each of the three 2-tori and $\alpha=x,y$ labelling the two coordinates of each torus, we have for type IIA:
\be
t_i=R_x^i R_y^i, \qquad s=e^{-\mfs}R^1_xR^2_xR^3_x, \qquad u_i=e^{-\mfs}R_x^iR_y^j R_y^k, \qquad i\neq j\neq k \neq i
\ee
and for type IIB:
\be
t_i=e^{-\mfs} R_x^jR_y^j R_x^k R_y^k, \qquad s=e^{-\mfs}, \qquad u_i=\frac{R_y^i}{R_x^i}, \qquad i\neq j\neq k \neq i
\ee
 
In each case if we fix $R_x^i=R_y^3=\ell $ and $R_y^1=R_y^2=L\gg \ell$ then the argument of the logarithm becomes $ s\left( \prod_{i=1}^2 t_i \right) \left( \prod_{i=1}^2 u_i \right) \propto L^6 \propto \cV^3$ as desired.  Although it is encouraging that multiple moduli can be in principle sufficiently large in this way, a full 10D provenance also requires a demonstration that this can be achieved in a concrete construction that stabilizes all moduli (which goes beyond the scope of this paper).
 
In order to have the no-scale structure on which Yoga-model success relies we would require any unfixed moduli to appear in the K\"ahler potential as $K = - 3 \ln F$ where $F$ is a homogeneous degree-one function under identical rescalings of all the moduli \cite{UVShadows, Burgess:2008ir}. It is simplest if this occurs with $F = \cT+\ol \cT$ depending only on a single field (as was chosen in \cite{YogaDE}), but it can also happen when more than one field is involved,\footnote{If several no-scale fields are involved homeopathic suppression of dilaton couplings would be required for all of them.} such as if $F=(stu)^{1/3}$ corresponding to the case \pref{STUK}.

No-scale moduli also cannot appear in (and so be fixed by) the superpotential or the $D$-term potential, leaving their energetics to be determined by the K\"ahler potential using RG stabilization as was done for $\tau$ in \cite{YogaDE}. Any other moduli are assumed to be stabilized in a supersymmetric way, such as by allowing them to appear in the superpotential. For instance, in the above example we might imagine that this is done so that $t_2$ and $u_2$ are stabilized supersymmetrically, with $\langle t_2 \rangle = \lambda_t = (M_s L)(M_s\ell)$ and $\langle u_2 \rangle = \lambda_u = L/\ell$. Then by rescaling the remaining unfixed fields by $\cT := \lambda_t \cT_1$ and $\cU := \lambda_u \cU_1$ the correct extra-dimensional shape is obtained if RG stabilization is chosen to ensure $\langle s \rangle\sim\langle t \rangle\sim \langle u\rangle$ are all order $\cV_2 \propto L^2$. Alternatively one could imagine supersymmetrically fixing all moduli except one ($\cT_1, say$), and redefining the remaining modulus by $\cT^3 \propto \cT_1$ with numerical coefficient chosen to contain the vevs of the fixed moduli: $\langle s \rangle$, $\langle t_2 \rangle$, $\langle u_i \rangle$.

If more than one modulus survives into the low-energy theory each would contain potentially dangerous matter couplings: in the above example these also would come with a K\"ahler potential of the form $K = - \ln[(\cS+\ol \cS)(\cT+\ol \cT)(\cU +\ol \cU) - k + \cdots]$ with $s = \cS+\ol \cS$, $t = \cT+\ol \cT$ and $u = \cU +\ol \cU$ to be RG-stabilized at size $L^2$ (we have verified that this structure preserves the low-energy Yoga-type suppressions found in \cite{YogaDE}). But because all Standard Model particles have masses proportional to $e^{-K/6}$, in such a framework all three of the fields $s$, $t$ and $u$ would couple to  Standard Model particles as Brans-Dicke scalars with large coupling constants. They also turn out to be very light, raising the threat that each could be ruled out by precision tests of GR within the solar system. 

Remarkably, however, each dilaton also comes with its own axionic partner and because the leading target-space metric is derived from $K = - \ln(stu)$ it is a product metric built from three independent copies of the $SL(2,R)$ invariant metric that leads to the kinetic terms given in \pref{SL2Rkin}. As a result the screening mechanism of \cite{Homeopathy} can in principle be applied to each multiplet separately, providing they can have the required axion-matter couplings, potentially allowing all three to evade detection in solar-system tests of GR. Although these more complicated multi-modulus examples might remain viable, in what follows we focus purely on the case where only a single axiodilaton pair survives at low energies.

\section{Axio-dilaton phenomenology above eV energies}
\label{sec:UVpheno}

The above picture provides the framework required for investigating physical questions involving energies above the eV scale. These include in particular constraints on axiodilaton properties that rely on the modelling of the interiors of stars and macroscopic sources, since these often involve environments hotter than eV scales. 

\subsection{Relevance to axion constraints}

In a nutshell, we have seen that the two model-independent $S$ and $T$ axions can couple to ordinary matter, with $S$ doing so with strength $1/f_a$ and $T$ doing so with gravitational strength $1/M_p$. Extra-dimensional gauge invariance also forbids a direct QCD-like coupling to the $T$-type axion. 

Because axion-matter couplings as strong as $1/f_a$ would have been detected, couplings of the form \pref{awedgeint} must be forbidden in the UV completion; any $S$-type axion found at low energies within a viable Yoga model must be an ALP and not directly couple to brane-localized Standard Model fields. Low-energy $T$-type axions are also predicted to be ALPs and are automatically photo-phobic, but can couple to other ordinary fields with strength $1/M_p$.

\begin{figure}[t]
\begin{center}
\includegraphics[width=77mm,height=55mm]{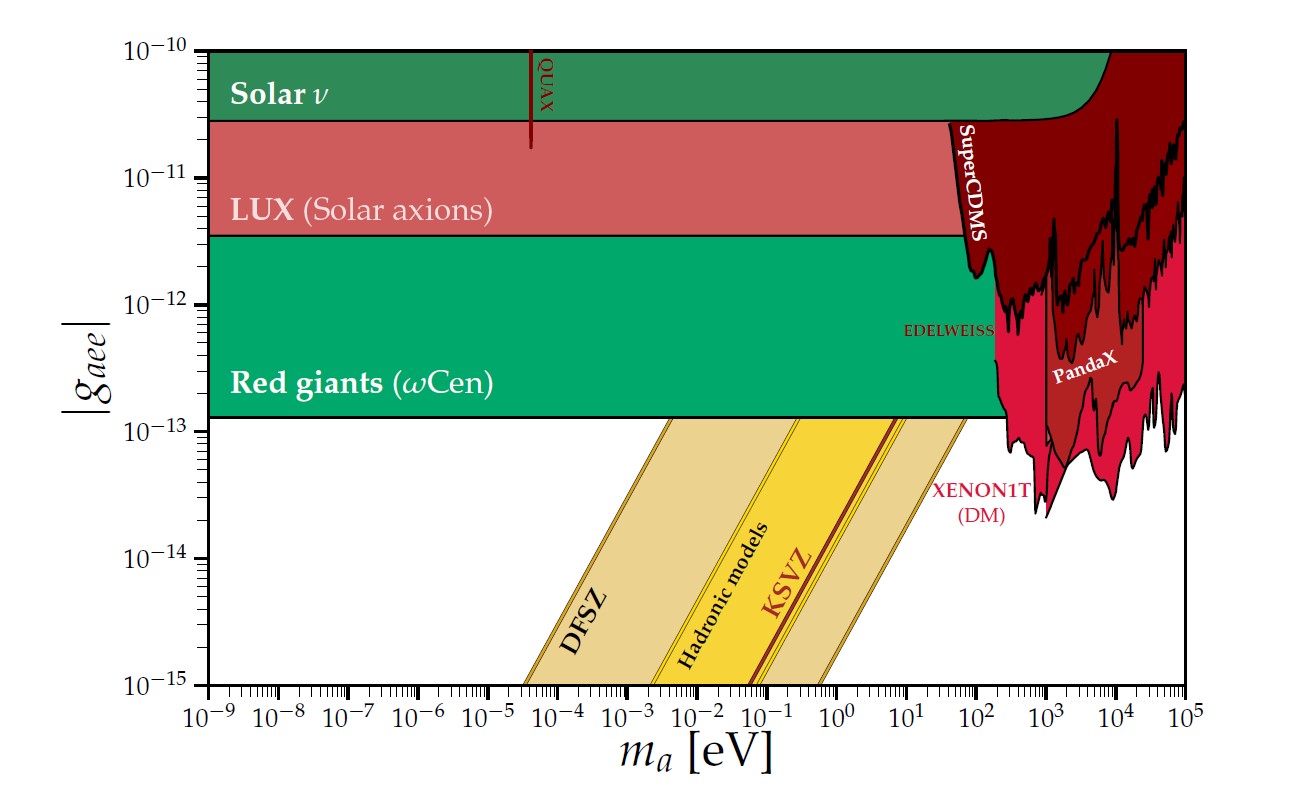} 
\includegraphics[width=77mm,height=55mm]{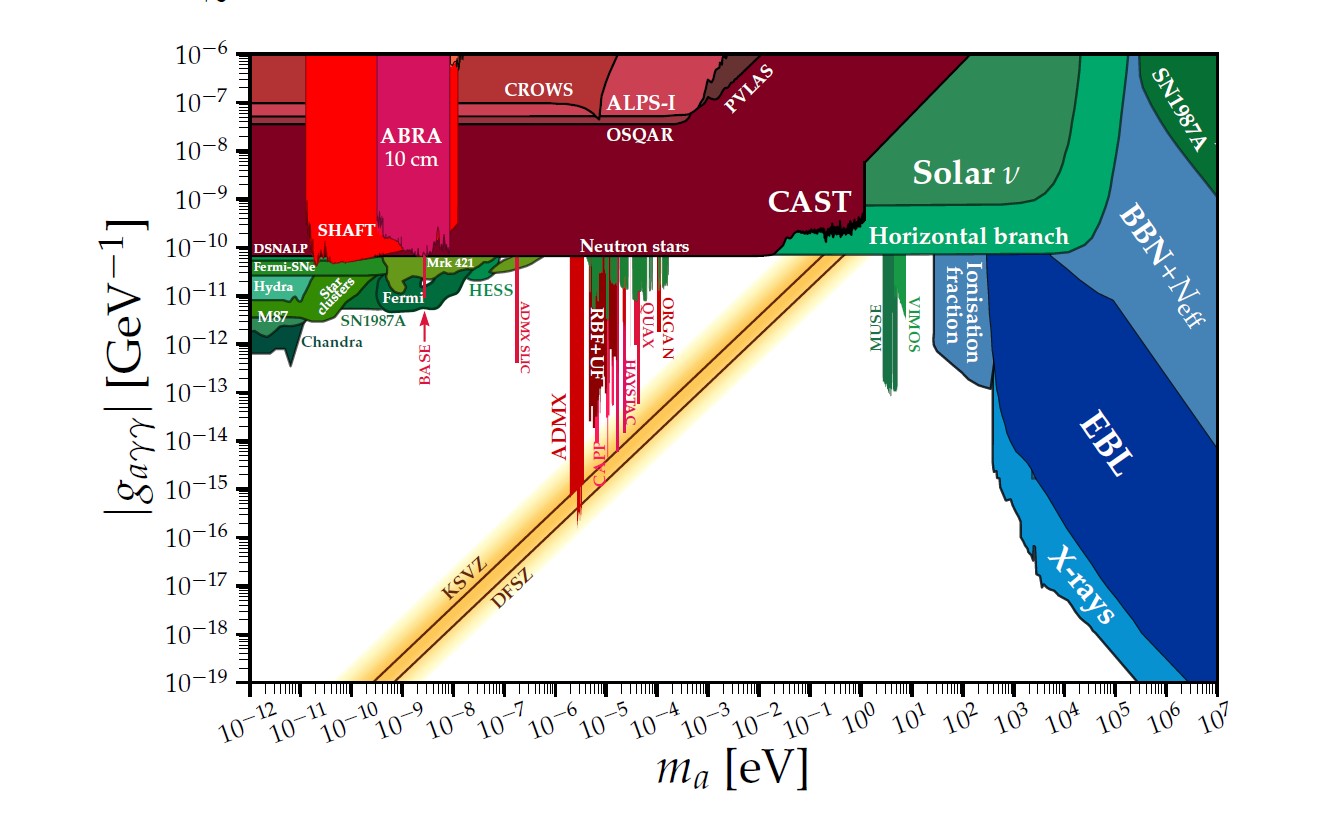} 
\caption{Constraints on axion-electron and axion-photon couplings from various astrophysical observations and lab searches \cite{PDG, gitFigs}.} \label{Fig:ElossBound} 
\end{center}
\end{figure}

\begin{figure}[h]
\begin{center}
\includegraphics[width=120mm,height=80mm]{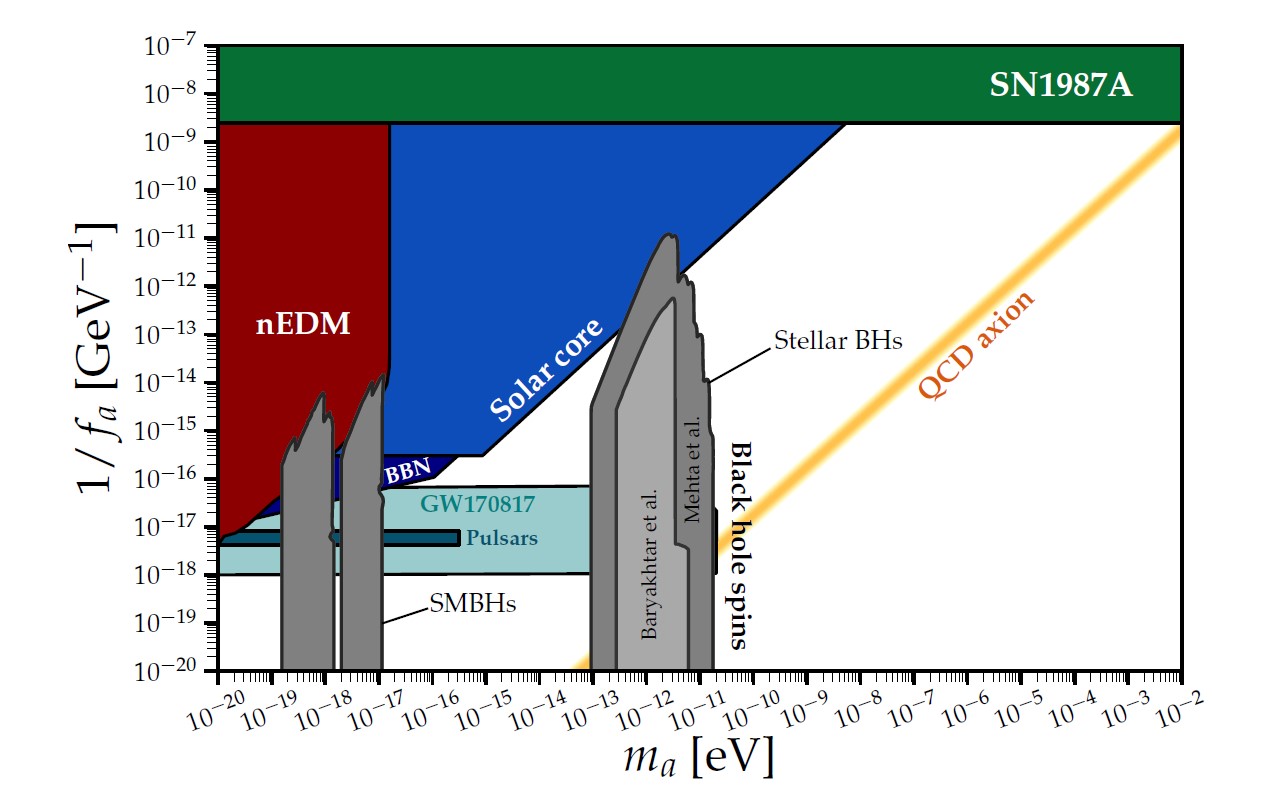} 
\caption{Constraints on axion decay constant vs mass, showing in particular the constraints on very light Planck-coupled axions derived in \cite{SuperradianceBounds, Hook:2017psm, Zhang:2021mks}. (Figure taken from \cite{PDG, gitFigs}.)} \label{Fig:FaxBound} 
\end{center}
\end{figure}

A compilation of constraints on the electron and photon couplings for such ALPs is shown in Fig.~\ref{Fig:ElossBound}, from which we also see that ALPs with gravitational strength $\cF^{-1} \sim M_p^{-1} \sim 10^{-18}$ GeV couplings are largely unconstrained. Fig.~\ref{Fig:FaxBound} provides several fairly strong constraints for gravitationally coupled axions coming from gravitational-wave and pulsar observations. The constraints coming from black hole super-radiance \cite{SuperradianceBounds} provide constraints that are largely independent of coupling strength $\cF^{-1}$ or $f_a^{-1}$, but do so only for a specific mass window. The constraints labelled `pulsars' and `GW170817' apply for a wide range of masses right down to $\cF^{-1} \sim 1/M_p$ \cite{Hook:2017psm, Zhang:2021mks}, but these rely more specifically on the existence of axion couplings to QCD and so would not directly apply for $S$-type or $T$-type ALPs. These must be revisited however should the light $S$- or $T$-type axions mix appreciably with another type of axion that does couple to QCD.

\subsection{Constraints with large dimensions}

Energy-loss bounds like the ones appearing in the Figures build on the observation that weakly coupled particles  provide an efficient way for astrophysical bodies to radiate away their energy. The bounds come from requiring this process not be so efficient that it would prevent such objects surviving long enough to have been detected with the properties they are observed have. 

The UV completion is essential to understanding these bounds because the energies involved (such as for red giants or supernovae) are typically much greater than the eV-sized KK scales associated with its onset. For a mode coupling strength $1/\cF$ with $\cF \sim M^2L \sim M_p$ the emission rate for any specific KK mode of energy $M \gg E \gg 1/L$ is $\Gamma_\ssE \sim (E^3/\cF^2) \sim (E^3/M_p^2)$ is at present not observable.

Energy-loss bounds are nevertheless important for extra dimensional models because of the enormous available phase space into which energy can be lost. These constrain the extra-dimensional UV scale $M$ because the density of states for momenta much higher than the KK scale is $\exd n/\exd^2k \sim L^{2}$ and so the total emission rate up to a maximum energy $M \gg Q \gg 1/L$ is of order 
\be
   \Gamma_{\rm tot} \sim \int_{E<Q} \exd^2k \left(\frac{\exd \Gamma}{\exd n} \right) \frac{\exd n}{\exd^2 k} \sim \int  _{E<Q} \exd^2 k \; L^2  \left( \frac{E^3}{F^2} \right) \sim \frac{Q^5}{M^4} \,.
\ee
Careful evaluation of the emission rate into extra dimensions in this way imposes significant constraints on the scale $M$ in the extra dimensions for large-dimension models (for which $Q \gg 1/L$ even at MeV scales) \cite{LED2, LEDBounds}, leading for supersymmetric extra dimensions to constraints of order\footnote{Indeed it is because these constraints were this high that that energy-loss signals of extra dimensions were unlikely in the $E < 14$ TeV collisions at the LHC.} $M \gsim 20$ TeV \cite{SLEDBounds}. Yoga models satisfy these constraints because for them the extra-dimensional scale is (by design) higher than this. 

\subsubsection*{Early-universe cosmology}

A well-known complication of large-dimensional models is that the gravitational response of cosmology to ordinary matter should be cast within an extra-dimensional framework right up until relatively recently. For extra dimensions at eV scales a 4D description along the lines given in the Yoga-model cosmologies described in \cite{YogaDE} should suffice for temperatures at and below an eV, ensuring that extra-dimensional UV completions play no role starting around recombination. 

In principle the physics of earlier epochs could require fully evolving within the 6D gravitational response of the UV theory, but this also might not be strictly necessary depending on the universe's earlier history. Unfortunately very little is known about cosmological solutions to the full 6D field equations (see however \cite{Goldilocks} for fully 6D explorations of inflationary dynamics). This also complicates fully analysing the origins of primordial fluctuations, and so in particular whether late-time isocurvature perturbations \cite{AxionIsocurvature} are necessarily present (whose observed absence \cite{Planck:2018jri} constrains theories with light scalars present at late times).

Broad circumstances under which cosmology might remain effectively 4-dimensional are considered in \cite{LED2}, who argue that complicating extra-dimensional effects (like energy loss into the extra dimensions) can conceivably be neglected up to temperatures above those relevant to Big Bang Nucleosynthesis (BBN). Care must also be taken to ensure that extra-dimensional KK modes cannot decay into SM degrees of freedom (such as photons) at late times, since doing this excessively can ruin the success of standard Hot Big Bang cosmology (for a review see \cite{Clifton:2011jh}). We here also assume such arguments apply, justifying use of a 4D framework.   

\subsection{Relationship to SLED models}
\label{ssec:SLEDrelation}

We see that Yoga models are very similar to earlier SLED models in their implications for UV (above eV scale) phenomenology. They resemble one another strongly because the details of the geometry of the two large extra dimensions are largely irrelevant for applications to energies much larger than an eV. For modes with energies $Q \gg 1/L$ wavelengths $\lambda \sim 1/Q$ are so short that an expansion in powers of $\lambda/L$ converges extremely quickly, and predictions can become indistinguishable from the $L \to \infty$ limit. 

It is the low-energy predictions that {\it do} depend more sensitively on extra-dimensional geometry, such as the mass spectrum of the lightest moduli and the energetics that generates their scalar potential (and stabilize their expectation values).  It is there that we should seek differences between Yoga and SLED models, and indeed we argue that it is the {\it absence} of the Yoga mechanism's no-scale structure and relaxation mechanisms that ultimately precluded SLED models from giving lower vacuum energies than they did. 

To see how this works it is useful to imagine a framework within which both Yoga and SLED models are particular limits. For these purposes it is useful to work within the concrete example of gauged chiral 6D supergravity \cite{NishinoSezgin}, within which SLED models explore vacua described by deformations of the explicit solutions of \cite{SalamSezgin} for which the presence of a background Maxwell field, $\langle F_{mn} \rangle$, stabilizes the extra dimensions into deformed 2-spheres (`rugby ball' geometries). The effective four-dimensional supergravity described by these solutions is as given in \cite{SalamSezgin4D}, and involves complex $S$ and $T$ moduli, much as described above in \S\ref{ssec:Unitarization} where $L$ is the sphere's radius modulus and $\mfs$ is the 6D dilaton. 

It is useful to think of the Salam-Sezgin solution to be a deformation of the toroidal compactifications that are more appropriate in the limit $\langle F_{mn} \rangle = 0$, in which we would expect there to be an additional complex modulus, $U$, associated with the toroidal complex structure. From the 4D point of view we should imagine that the energy associated with the nonzero Maxwell flux introduces a dependence of the superpotential on $U$ in such a way that $\langle U \rangle$ can be fixed at a supersymmetric mimimum, found by solving $D_\ssU W = 0$. The Maxwell flux also introduces a Fayet-Iliopoulos term that stabilizes $T$ at a supersymmetric point \cite{SalamSezgin4D}, leaving only $S$ free to parameterize a flat direction along which $W = 0$ and so supersymmetry remains unbroken. The low-energy 4D supergravity describing this last flat direction is {\it not} a no-scale model. 

For the Yoga mechanism we might instead choose to leave all three moduli unfixed by $W$ and by $D$-term potentials, so that they can provide a nonsupersymmetric no-scale limit whose flat direction is lifted (and moduli stabilized at much larger values) by the RG mechanism described in \cite{YogaDE}. The simplest 2-torus -- with two similarly large sides $L_1 \sim L_2 \sim L$ -- predicts $S , T \propto L^2$ and $U \sim \cO(1)$ and so gives $\tau$ (identified by equating \pref{KSS} to $-3\ln \tau$) is too small, but this can be alleviated using constructions along the lines described in \S\ref{ssec:AsymmetricYC}. 

This perspective shows why the Salam-Sezgin based SLED story never in practice led to a sufficiently small vacuum energy \cite{SLEDNoGo} and why the Yoga model does better. But it also shows why the successes of the SLED story in accounting for the small size of UV quantum corrections \cite{UVSLEDb, UVSLEDBulk} should also apply to Yoga models: vacuum energies on the brane act to curve the extra dimensions and not the dimension seen in cosmology, and supersymmetry suppresses UV contributions from the bulk. Yet the localization of SM particles on the brane also explains why many 4D inferences about their properties ({\it e.g.}~$\tau$-dependence of masses) remains reliable despite the presence of extra dimensions. In this sense Yoga models capture the best of both the 4D and 6D perspectives.

\section{Long-range forces: shift-symmetric vacuum}
\label{sec:IRscreening}

We now change gear and use the above insights into axiodilaton-matter couplings to examine more closely the implications of the above types of axio-dilaton/matter couplings for very low-energy phenomenology. In particular we focus on the parameter regime for which $V_{\rm min} \lsim 10^{-80}M_p^4$, for which the dilaton Compton wavelength exceeds 10 km and so bounds on gravitational strength Brans-Dicke forces become a problem, clarifying the role of axion shift symmetry and the low-energy role of the relaxon field $\chi$. 

Our interest in particular is to use the knowledge of matter-axion couplings to model the interior solutions for gravitating sources, and see whether the bound-evading exterior solutions of \cite{Homeopathy} are actually generated.  The discussion proceeds differently depending on whether or not the axion shift symmetry is broken by the vacuum so we consider these two options separately, restricting in this section to the case where the vacuum preserves the $SL(2,R)$ invariance of the leading order axiodilaton field equations.

\subsection{New exterior solutions}
\label{ssec:HomeoAxCouplings}

We start by constructing a new broad class of solutions to the axiodilaton field equations, extending the spherically symmetric solutions found in \cite{Homeopathy} to include a broad class of solutions with less symmetry. To this end we explore the semiclassical implications of the action
\be \label{axiodilL}
   \cL  
    = - \sqrt{-g} \; M_p^2 \left[ \frac{\cR}2 + \frac34  \left( \frac{\partial^\mu \tau \, \partial_\mu \tau + \partial^\mu \mfa \, \partial_\mu \mfa}{\tau^2} \right) \right] + \cL_m \,,  
\ee
where $\cL_m = \cL_m(\tilde g_{\mu\nu},\mfa,\psi)$ is the matter lagrangian density, in which $\psi$ is a generic matter field and $\tilde g_{\mu\nu} = A^2(\tau) \, g_{\mu\nu}$ with $A = \tau^{-1/2}$. The corresponding axiodilaton field equations are
\be \label{AxDilEOM}
      \Box \tau  - \frac{1}{\tau} \Bigl( \partial_\mu \tau \, \partial^\mu \tau - \partial_\mu \mfa \, \partial^\mu \mfa \Bigr) - \frac{\tau \, T }{3 M_p^2} = 0 \,, \qquad
      \Box \mfa  - \frac{2}{\tau} \,\partial_\mu \tau \, \partial^\mu \mfa + \frac{\tau^2  \cA}{3 M_p^2}  = 0 \,,
\ee
and
\be \label{TTEinstMAX}
  \cR_{\mu\nu} + \frac{3}{2\tau^2} \Bigl(  \partial_\mu \tau \, \partial_\nu \tau +  \partial_\mu \mfa \, \partial_\nu \mfa \Bigr) + \frac{1}{M_p^2} \left( T_{\mu\nu} - \frac{T}2  \, g_{\mu\nu} \right) = 0 \,,
\ee
where we write $T = g_{\mu\nu}T^{\mu\nu}$ with
\be
   T^{\mu\nu} := \frac{2}{\sqrt{-g}} \left( \frac{\delta S_m}{\delta g_{\mu\nu}} \right)_{\psi,\mfa} \quad \hbox{and} \quad \cA := \frac{2}{\sqrt{-g}} \left( \frac{\delta S_m}{\delta \mfa} \right)_{\psi,\tilde g_{\mu\nu}}\,.
\ee

As is easy to verify directly, the field equations are equivalent to divergence relations 
\be \label{JEOMs}
   D_\mu J^\mu_{(a)} 
   = - \frac{\cA}{3M_p^2} \,, \quad
     D_\mu J_{(s)}^\mu 
     =  \frac{(T - \mfa \,\cA)}{3M_p^2} \quad \hbox{and} \quad
     D_\mu J_{(n)}^\mu 
     = \frac{(\mfa^2 - \tau^2) \cA - 2\mfa \,T}{3M_p^2} \,,
\ee
for the following three currents 
\be \label{Jdefs}
  J_{(a)}^\mu = \frac{\partial^\mu \mfa}{\tau^2} \,, \quad
   J_{(s)}^\mu = \frac{\partial^\mu \tau}{\tau} + \frac{\mfa \,\partial^\mu \mfa}{\tau^2} \quad \hbox{and} \quad
   J_{(n)}^\mu = \frac{(\tau^2 - \mfa^2)}{\tau^2} \; \partial^\mu \mfa - \frac{2\mfa}{\tau} \; \partial^\mu \tau  \,.
\ee
In particular these are all conserved in regions where $T^{\mu\nu} = \cA = 0$, reflecting the classical $SL(2,R)$ invariance enjoyed by the axio-dilaton part of the action \pref{axiodilL}. 

Previous sections tell us that the leading microscopic axion/matter couplings have the form
\be \label{Lam1}
   \cL_{m} \ni - \sqrt{-g} \; \partial_\mu \mfa \, \cJ^\mu
\ee
with currents $\cJ^\mu$ as given in \pref{Jpieces} and \pref{JpiecesA}. This implies $\cA = 2 D_\mu \cJ^\mu$ and so the first of \pref{JEOMs} implies the combination $ \mfJ^\mu := J_{(a)}^\mu + \frac23(\cJ^\mu/M_p^2)$ satisfies 
\be \label{Axcurrmod}
   D_\mu \mfJ^\mu = D_\mu\left( J_{(a)}^\mu + \frac{2\cJ^\mu}{3M_p^2} \right) = 0 
\ee
even in the presence of matter. This is Noether's theorem at work because \pref{Lam1} preserves the invariance under constant axion shifts: $\mfa \to \mfa + c$. 

For the Chern-Simons interactions given in \pref{JpiecesA} the current $\cJ^\mu$ in \pref{Lam1} is not gauge invariant, although $D_\mu \cJ^\mu$ is (and is nonzero). In this case gauge invariance is more transparent if we write $\cL_{am}$ in the form 
\be
   \cL_{am} =  \sqrt{-g} \;  \mfa \, D_\mu \cJ^\mu \,.
\ee
Broken shift symmetry then obstructs finding a gauge-invariant current $\mfJ^\mu$ satisfying \pref{Axcurrmod}. 

\subsubsection*{General class of exterior solutions}

We next derive a broad class of exact solutions to the classical scalar equations exterior to a source that extend the weak-field spherically symmetric ones found in \cite{Homeopathy} and the strong-field spherically symmetric solutions of \cite{StrongSolns} (see also \cite{choi} for an independent discussion) .

We first notice that the currents \pref{Jdefs} imply that for any constant $\alpha$
\be \label{semicirclediff}
   J_{(s)}^\mu - \alpha J_{(a)}^\mu  =  \frac{ \partial^\mu[ \tau^2 + (\mfa - \alpha)^2]}{2\tau^2}  \,,
\ee
reflecting the important role played in \cite{Homeopathy} by semicircles $\tau^2 + (\mfa - \alpha)^2 = \beta^2$ that are the geodesics of the target space metric $\exd s^2 = (\exd \tau^2 + \exd \mfa^2)/\tau^2$. This suggests the ansatz 
\be \label{semicircleAnsatz}
  \tau =\frac{\beta}{\cosh X} \quad \hbox{and} \quad
  \mfa = \alpha + \beta \tanh X \,,
\ee
with $\alpha$ and $\beta$ constants and $X=X(x)$ a general function of position and time. This ansatz implies the axion current becomes
\be \label{JAvsX}
  J_{(a)}^\mu = \frac{\partial^\mu \mfa}{\tau^2} = \frac{\partial^\mu X}{\beta} 
  \quad \hbox{and so} \quad
  \partial_\mu \Bigl( \sqrt{-g} J_{(a)}^\mu \Bigr) = \frac{\sqrt{-g} \; \Box X}{\beta} \,,
\ee
which shows $\Box X = 0$ suffices to guarantee conservation of $J_{(a)}^\mu$ (as \pref{JEOMs} shows is true whenever $\cA=0$).  But \pref{semicirclediff} and \pref{semicircleAnsatz} together imply $J_{(s)}^\mu = \alpha \, J_{(a)}^\mu$ so $J_{(s)}^\mu$ is also conserved. Similarly
\be
  J_{(n)}^\mu 
  = \frac{\partial^\mu X}{\beta} \left[ \frac{\beta^2}{\cosh^2 X} - (\alpha + \beta \tanh X)^2 + 2\beta (\alpha + \beta \tanh X) \tanh X \right] 
  = \left( \frac{\beta^2 - \alpha^2}{\beta} \right) \partial^\mu X \,,
\ee
and so $\Box X = 0$ also suffices to ensure conservation of $J_{(n)}^\mu$ (as must hold whenever $T = \cA = 0$).  

It follows that the ansatz \pref{semicircleAnsatz} promotes any solution of $\Box X = 0$ to a solution of the axio-dilaton equations \pref{AxDilEOM} exterior to sources (where $\cA = T_{\mu\nu} = 0$). All that is required to obtain a full solution then is to solve the Einstein equations \pref{TTEinstMAX}. Using
\be 
  \partial \tau = - \frac{\beta \sinh X}{\cosh^2X} \; \partial X \quad \hbox{and} \quad
  \partial \mfa = - \frac{\beta}{\cosh^2 X} \; \partial X
\ee
allows the derivation of the identity $\tau^{-2} ( \partial_\mu \tau \partial_\nu \tau + \partial_\mu \mfa \, \partial_\nu \mfa ) = \partial_\mu X \, \partial_\nu X$, and so \pref{TTEinstMAX} becomes
\be \label{TTEinstMAXX}
  \cR_{\mu\nu} + \frac{3}{2} \;  \partial_\mu X \, \partial_\nu X + \frac{1}{M_p^2} \left[ T_{\mu\nu} - \frac12 \, g^{\lambda\rho} T_{\lambda\rho} \, g_{\mu\nu} \right] = 0 \,.
\ee
In the absence of sources we can therefore take {\it any} solution of the coupled Einstein/Klein-Gordon equations and promote it into a solution of the full Einstein-axiodilaton equations. 

\subsubsection*{Multipole solutions}

Specializing to static solutions on a flat spacetime metric, in spherical coordinates $(r, \theta, \xi)$ the general solution to $\nabla^2 X = 0$ is
\be
    X(r,\theta,\xi) = \sum_{\ell =0}^\infty \sum_{m=-\ell}^\ell \left( a_{\ell m} r^\ell + \frac{b_{\ell m}}{r^{\ell +1}} \right) Y_{\ell m}(\theta, \xi) 
\ee
for spherical harmonics $Y_{\ell m}(\theta,\xi)$ and arbitrary constants $a_{\ell m}$ and $b_{\ell m}$. Requiring $\tau$ to be finite and nonzero at spatial infinity then implies $a_{\ell m} = 0$ for all $\ell \neq 0$. Further assuming axial symmetry (independence of $\xi$) then sets all coefficients to zero when $m \neq 0$, leading to a standard multipole form
\be \label{DipoleXForm}
    X(r,\theta) 
    = a_0 + \sum_{\ell = 0}^\infty  \frac{b_{\ell}}{r^{\ell +1}} \; P_{\ell}(\cos\theta) \,.
\ee

The integration constants appearing in this solution can be read off from the radial components of the currents just outside the source, as usual. For instance the radial flux of the current $J_{(a)}^\mu$ at $r = R$ is given -- {\it c.f.} \pref{JAvsX}  -- by
\be
  R^2  J_{(a)}^r(R,\theta) =R^2 \left( \frac{\partial_r X}{\beta} \right)_{r = R} =- \sum_{\ell = 0}^\infty  (\ell+1)\frac{b_{\ell}}{\beta R^{\ell}} \; P_{\ell}(\cos\theta)  \,,
\ee
from which all of the $b_\ell$'s can be read off by equating the $\theta$-dependence of both sides. 

Spherical symmetry corresponds to $\ell = 0$, for which the above agrees with the spherically symmetric solutions given in \cite{Homeopathy} if we define $b_0 = \beta \gamma$ and $a_0 = \delta$ so that $X = \delta + (\beta \gamma/r)$ and the leading far-field behaviour is given by
\be  \label{LargerAsympt1}
   \tau  =  \frac{\beta}{\cosh X} = \tau_\infty \left[ 1 - \frac{\beta \gamma}{r} \, \tanh \delta + \cdots \right] \quad\hbox{and} \quad
    \mfa= \mfa_\infty  - \frac{\beta^2\gamma}{r \cosh^2\delta} + \cdots  \,.
\ee

\subsubsection*{Test-particle motion}

The above solutions determine the motion of test particles, which for weak axion/matter couplings move along geodesics of the Jordan-frame metric $\tilde g_{\mu\nu} = A^2 g_{\mu\nu} = g_{\mu\nu}\sqrt{\tau_\infty/\tau}$. Constraints coming from measurements of test-particle motion are therefore conveniently described by parameterizing this metric in terms of post-Newtonian parameters. 

For the above spherically symmetric solutions the relevant PPN parameters turn out to be \cite{Homeopathy}
\be \label{yogagammaPPN}
  \gamma_\PPN   =  \frac{1 - 2 \lambda_{\rm eff} \tanh \delta}{1 + 2 \lambda_{\rm eff} \tanh \delta} \,,
\ee
and\footnote{In principle $\beta_\PPN$ is sensitive to $1/r^2$ contributions to $\tilde g_{tt}$, which in PPN formulations is written $\tilde g_{tt} = -1 + 2U - 2\beta_\PPN U^2 + \cdots$ where $U$ is the Newtonian potential. We quote here the prediction of {\cite{Homeopathy}} for monopole sources because in practice a dipole contribution to the dilaton coming from the $\ell = 1$ term of {\pref{DipoleXForm}} is indistinguishable from a very small but nonzero dipole moment for the gravitating source in $U$.}
\be \label{yogabetaPPN}
  \beta_\PPN  
   = \frac{1 + 4 \lambda_{\rm eff} \tanh \delta + 4 \lambda_{\rm eff}^2}{(1 + 2 \lambda_{\rm eff} \tanh \delta)^2} = 1 + \frac{4 \lambda_{\rm eff}^2}{(\cosh\delta + 2 \lambda_{\rm eff} \sinh\delta)^2}  \,,
\ee
where $\lambda_{\rm eff}$ is defined by
\be \label{BDeffDef}
   \beta\gamma =: -4 \lambda_{\rm eff} GM 
\ee
with $M = \int \exd^3x \; \rho$ (for Einstein-frame energy density $\rho \simeq -T$) being the leading Newtonian contribution to the gravitational (and inertial) mass in Einstein frame. Agreement with solar-system tests requires both $|\beta_\PPN-1| \lsim 10^{-5}$ and $|\gamma_\PPN -1| \lsim 10^{-4}$ and these combined tell us that $\lambda_{\rm eff} \lsim 10^{-5}$ largely independent of the value of $\delta$.

Brans-Dicke theory \cite{Jordan, BransDicke, Dicke:1964pna, Brans} can be defined as a massless and canonically normalized scalar field $\mfs_\BD$ that couples minimally to gravity in Einstein frame but couples to matter through the metric $\tilde g_{\mu\nu} = A^2 g_{\mu\nu}$ with $A =\exp(\mfg \,\phi_\BD/M_p)$. The parameter $\mfg$ defines the Brans-Dicke coupling.\footnote{In our conventions $\mfg$ is related to the conventional Brans-Dicke parameter $\omega$ by $2\mfg^2 = (3 + 2\omega)^{-1}$.} Using $A = \tau^{-1/2}$ and \pref{axiodilL} reveals $\ln\tau$ to be a Brans-Dicke scalar (at least in the absence of the axion, to leading order in $1/\tau$) with coupling $\mfg^2 = \frac16$. 

Eqs.~\pref{yogagammaPPN} and \pref{yogabetaPPN} go over to standard Brans-Dicke results in the limit $\delta \to \infty$ and $\lambda_{\rm eff} \to \mfg^2$, but the observation made in \cite{Homeopathy} is that there is no {\it a priori} reason $\lambda_{\rm eff}$ and $\delta$ obtained from the solution exterior to a source need agree with $\mfg^2 = \frac16$ and $\delta \to \infty$ when both axion and dilaton are present. They in general differ because $\lambda_{\rm eff}$ is given in terms of the integration constants $\beta$ and $\gamma$ by \pref{BDeffDef} and these (and $\delta$) must be obtained by matching to an interior solution at $r = R$, just outside the source's surface. For instance, integrating the $D_\mu J_{(a)}^\mu$ and $D_\mu J_{(s)}^\mu$ equations of \pref{JEOMs} through the interior of the source leads to
\be \label{Aconsradint2}
   \gamma =  R^2 \left( \frac{\partial_r\mfa}{\tau^2} \right)_{r=R} = - \frac{1}{3M_p^2}\int_0^R \exd r \; r^2 \cA(r)   \,,
\ee
and
\be \label{Sconsradint2}
   \gamma \alpha = R^2 \left( \frac{\partial_r\tau}{\tau} + \frac{\mfa \, \partial_r\mfa}{\tau^2}\right)_{r=R} = - \frac{1}{3M_p^2} \int_0^R \exd r\; r^2 \Bigl[\rho(r) + \mfa(r) \,\cA(r) \Bigr] 
   \,,
\ee
while $\beta^2 = \tau^2(R) + [\mfa(R)- \alpha]^2$. An equivalent (and often more useful) way to rewrite \pref{Sconsradint2} combines it with \pref{Aconsradint2} to give
\be \label{Sconsradint3}
   \gamma \Bigl( \mfa(R) - \alpha \Bigr) =-  R^2 \left( \frac{\partial_r\tau}{\tau}  \right)_{r=R} = \frac{1}{3M_p^2} \int_0^R \exd r\; r^2 \Bigl[\rho(r) + \Bigl( \mfa(r) - \mfa(R) \Bigr) \cA(r) \Bigr] 
   \,.
\ee

Of course just because a theory has a potential escape route from dangerous observations doesn't mean that it necessarily uses it. Saying more about how Yoga models fare once compared to {\it e.g.}~solar system tests requires knowing more about $\cA$, and so partially relies on the UV completion above eV scales provided here since these are required to formulate matter-axion interactions within astrophysical environments. We pursue interior solutions further in \S\ref{ssec:interiors} but first pause to describe constraints/opportunities that require only the solutions given above and in \cite{YogaDE}, to do with observable effects associated with $\tau$ providing position-dependent masses to ordinary particles. 

\subsection{Mass variation}

A signature prediction of these models is that all non-neutrino SM particle masses are proportional to $\tau^{-1/2}$, while neutrino masses plausibly vary like $\tau^{-1}$. Any spatial or temporal variation of $\tau$ implies a similar variation of these masses relative to the Planck scale (which in Einstein frame is fixed). Because all masses scale in the same way all non-neutrino mass {\it ratios} remain independent of spacetime position. 

We here estimate the magnitude of these mass variations, within the solar system and in cosmology. For bodies within the solar system we use the above explicit external solutions in the weak-field limit, since these anchor the dilaton profile for all bodies to a common reference value $\tau_\infty$ at spatial infinity. 

\subsubsection*{Variation between the surfaces of different celestial objects}

Consider first the variation between particle masses on the surface of different celestial objects, such as by comparing spectral lines on the surface of the Sun and on the surface of the Earth. Using $\beta\gamma = - 4   \lambda_{\rm eff} GM$ the external dilaton solution given in \pref{LargerAsympt1} becomes 
\be
  \frac{1}{ \tau(r)} =
  \frac{1}{\beta} \cosh \left[ \frac{4 \lambda_{\rm eff} GM}{r} - \delta \right] \simeq \frac{1}{\tau_\infty} \left[ 1 - \frac{4 \lambda_{\rm eff}GM}{r} \, \tanh \delta+\cdots \right] \,,
\ee
and so the change in a particle's mass between the surfaces of the sun and the earth (say) would be
\be
 \frac{m(R_\odot)-m(R_\oplus)}{m(R_\oplus)} =  \sqrt{\frac{\tau(R_\oplus)}{\tau(R_\odot)} } - 1 \simeq 2\lambda_{\rm eff} \left[ \left( \frac{GM}{R}\right)_\oplus - \left( \frac{GM}{R} \right)_\odot \right]\tanh\delta  \,.
\ee
Given that $(GM/R)_\oplus \sim 10^{-10}$ and $(GM/R)_\odot \sim 10^{-6}$ and solar system tests require $\lambda_{\rm eff} \lsim 10^{-5}$, we see that masses on the surface of the Sun are smaller than those on the Earth at most by about 1 part in $10^{11}$. It seems unlikely that spectral lines on the solar surface will soon be measured with sufficient accuracy to test this. 

\subsubsection*{Variation with altitude on Earth}

Mass variation with position near the Earth's surface is more likely to be testable (given the current precision of atomic clocks being 1 part in $10^{16}$). An estimate of the mass difference due to a change of height $h$ above the Earth's surface is
\be
 \frac{m(R+h)-m(R)}{m(R)} =  \sqrt{\frac{\tau(R)}{\tau(R+h)} } - 1 
 \simeq \frac{2\lambda_{\rm eff}h}{R} \left( \frac{GM}{R}  \right) \tanh \delta  \,.
\ee
This is at most order $10^{-18}$ for an altitude $h/R_\oplus \sim 10^{-3}$ above sea level on Earth (the altitude of Colorado, say) given that $(GM/R)_\oplus \sim 10^{-10}$. Although beyond the reach of current atomic clocks, it might become measurable in the not too distant future if $\lambda_{\rm eff} \tanh\delta$ is close to its upper bound. 

\subsubsection*{Variation inside gravitating bodies}

An estimate of the variation between masses at the center and surface of a gravitating body cannot be done using only exterior solutions, and requires solving the field equations in the presence of nonzero energy and axion-source density. A conservative estimate for the size of these mass variations interior to the Earth is obtained by assuming that it is unsuppressed by factors of $\lambda_{\rm eff}$ and is similar to the naive Brans-Dicke result:
\be \label{BDInternal}
 \frac{m(R)-m_0}{m_0} =  \sqrt{\frac{\tau_0}{\tau(R)} } - 1= \cO\left( \frac{GM}{R} \right)
\ee
which is order $10^{-10}$ (or $10^{-6}$) between the Earth's (or Sun's) centre and surface, for example. 

Although it seems unlikely to be able to determine ordinary particle masses accurately enough at the centre of the Earth (or other bodies) to test \pref{BDInternal}, the fact that neutrino masses can scale differently with $\tau$ than masses of ordinary matter means that this variation might conceivably alter the details of matter-dependent neutrino oscillations within the Sun or Earth. This might be hoped to be relevant for resonant oscillations, since resonance requires small neutrino mass differences, $\Delta m_\nu^2$, to coincide with equally small matter-induced energies, $G_\ssF n_e E_\nu$. Unfortunately although this could change the depth at which a resonance occurs it is unlikely to remove a resonance entirely, making it difficult to observe.

\subsection{Interior solutions and matching}
\label{ssec:interiors}

The simplest case assumes the vacuum preserves both the $SL(2,R)$ and shift symmetries, leaving open whether the couplings to matter also do so.  In this case the equations of motion governing the fields outside the source are those described above -- {\it c.f.}~eqs.~\pref{AxDilEOM} and \pref{TTEinstMAX} and the open question is how axiodilaton-matter couplings within a source's interior source these solutions. 

\subsubsection*{Brans-Dicke limit}

We start by examining more carefully the Brans-Dicke limit $\cA \to 0$ in which the source does not couple at all to the axion. Denoting radial derivatives by primes, the $\cA \to 0$ limit of \pref{Aconsradint2} implies  
\be \label{Aconsradint2y}
   \gamma =  R^2 \left( \frac{\mfa'}{\tau^2} \right)_{r=R} = - \frac{1}{3M_p^2}\int_0^R \exd r \; r^2 \cA(r)  \to 0 \,,
\ee
while \pref{Sconsradint3} instead says  
\be \label{Sconsradint3y}
   \gamma \Bigl( \mfa(R) - \alpha \Bigr) =-  R^2 \left( \frac{\tau'}{\tau}  \right)_{r=R} \to \frac{1}{3M_p^2} \int_0^R \exd r\; r^2 \, \rho(r)  =   \frac{2GM}{3} \,,
\ee
with the last equality defining $M$. Because the right-hand side is fixed this implies $\mfa(R) - \alpha$ must diverge as $\gamma \to \infty$. 

But $\lambda_{\rm eff}$ is determined from \pref{BDeffDef} by the product $\beta \gamma$ where $\beta^2 = \tau^2(R) + [\mfa(R)- \alpha]^2$, and so 
\be \label{geffresult}
   \lambda_{\rm eff} = -\frac{\beta\gamma}{4GM}  =  \frac{1}{4GM} \Bigl[\gamma^2  \tau^2(R) + \gamma^2 [\mfa(R)- \alpha]^2 \Bigr]^{1/2} \to  \frac{1}{6}  
\ee
as expected, after using \pref{Sconsradint3y} and assuming 
$\gamma\, \tau(R) \to 0$. The question is whether the $\cA$-dependence of \pref{Aconsradint2} and \pref{Sconsradint3} can reduce this result.

\subsubsection*{Unbroken shift symmetry inside the source}
\label{ssec:unbrokenshift}

Now comes a key observation: the constant $\gamma$ vanishes for spherically symmetric configurations if the axion/matter couplings inside the source preserve the axion's shift symmetry. To see why, recall from \S\ref{ssec:HomeoAxCouplings} that $\cA = 2 D_\mu \cJ^\mu$ when linear axion-matter couplings are shift-symmetric, which for spherically symmetric configurations on flat space implies $r^2 \cA = (r^2 \cJ)'$ for $\cJ = 2 \cJ^r$. The conservation law \pref{Axcurrmod} in this case then states that the radial flux of the current $\mfJ^\mu$ is $r$-independent:
\be \label{Aconsradm}
    \Bigl( r^2 \mfJ^r \Bigr)' = \left[ r^2 \left( \frac{\mfa'}{\tau^2} + \frac{\cJ}{3M_p^2} \right) \right]' = 0 
\ee
both inside {\it and} outside the source. But this means the radial flux must vanish because boundedness of the current implies $r^2 \mfJ^r \to 0$ as $r \to 0$ deep within the interior. But $\gamma = r^2 \mfJ^r = r^2 \mfa'/\tau^2$ for $r > R$ (outside the source) so it follows that $\gamma$ must also strictly vanish.   

Of course this does not mean that there is no axion field external to a source that couples to the axion in a shift-symmetric way. What it means is that shift-symmetric couplings generate higher multipole moments than the monopole, and so are not spherically symmetric. Although this means that the exterior solution obtained by matching from the interior has nonzero constants like $b_\ell$ in {\it e.g.}~\pref{DipoleXForm}, the key `monopole' constant $\gamma$ vanishes. Because of this the exterior scalar fields fall off too quickly in powers of $1/r$ to affect the prediction for the PPN parameters, and consequently cannot help suppress predictions relative to the Brans-Dicke result. 

This puts a premium on axion-matter couplings that break the shift symmetry, something that only happen for $\Phi$-type axions, at least withing the candidate interactions in the UV completions\footnote{$\Phi$ type axions can acquire QCD-like couplings through for example $\int B \wedge F \wedge F$ on-brane interactions, such as can arise in anomaly cancellation in higher dimensions.} described in \S\ref{sec:UVcompletion}. In principle such axions can also mix with the lower-mass counterparts, so in what follows we pursue the possibility that the axiodilaton could have shift symmetry breaking couplings to matter in order to see what this might imply for dilaton couplings. We first consider the case where matter couplings within a gravitating source break shift symmetry without the vacuum outside the source also doing so -- something that would {\it not} be appropriate limit if shift symmetry is broken by the QCD anomaly, but which can arise once the axion potential is subject to the relaxon dynamics in Yoga models (whose full discussion we are exploring but whose discussion we defer to future work). 


\subsubsection*{Broken shift symmetry within the source}
\label{ssec:brokenshift}

We next examine the case where the axion shift symmetry is broken by the matter-axion couplings but is not broken in the vacuum. 

For concreteness we assume the axion source density to be proportional to the energy density
\be\label{varepsdef}
   \cA = \varepsilon \, \rho
\ee
for a coefficient $\varepsilon$ that is itself in general axion-dependent; often a periodic function of $\mfb$ should a discrete subgroup of shifts remain unbroken. If the axion is a pseudoscalar then CP conservation would predict $\varepsilon(-\mfa) = - \varepsilon(\mfa)$ and so in particular $\varepsilon =  0$ when $\mfa = 0$ for any macroscopic object \cite{Wilczek}. This implies any axionic charge is necessarily suppressed by CP-violating couplings, and so $\varepsilon  \sim 10^{-17} - 10^{-19}$ if it is the Standard Model that provides the CP violation \cite{GeorgiRandallKhrip}. Although $\varepsilon$ can be larger than this if new physics provides the CP violation the experimental absence of a neutron electric dipole moment make it likely that $\varepsilon \lsim 10^{-10}$ \cite{EDMChiral}. 

The generic problem in this framework is to find $\varepsilon$ such that $|\lambda_{\rm eff}|$ is much smaller than its Brans-Dicke value of $\frac16$. What makes this difficult is \pref{geffresult}, which implies 
\be
  \Bigl|\lambda_{\rm eff}\Bigr| \geq \frac{|\gamma[\mfa(R) - \alpha]|}{4GM} \,, 
\ee
together with \pref{Sconsradint3}, which with \pref{varepsdef} implies
\be \label{Sconsradint3x}
   \gamma \Bigl( \mfa(R) - \alpha \Bigr) = \frac{8\pi G}{3} \int_0^R \exd r\; r^2 \rho(r) \Bigl[1 + \Bigl( \mfa(r) - \mfa(R) \Bigr) \varepsilon(r) \Bigr]    \,.
\ee
Small values for $\lambda_{\rm eff}$ require a significant cancellation within the square bracket of \pref{Sconsradint3x}, which is difficult for several reasons:
\begin{itemize}
\item First, \pref{Sconsradint3x} shows that $|\lambda_{\rm eff}| \geq \frac16$ whenever $[\mfa(r) - \mfa(R)] \varepsilon(r)$ is positive. When this is true the presence of the axion makes the source-dilaton coupling relevant to tests of gravity larger rather than smaller.
\item 
Second, it can be shown that $[\mfa(r) - \mfa(R)] \varepsilon(r) \geq 0$ whenever $\varepsilon(r)$ has a definite sign (positive or negative) because \pref{Aconsradint2} and the boundary condition $\mfa'(0) = 0$ imply $\mfa(r) - \mfa(R)$ then has the same sign as $\varepsilon(r)$. This observation is borne out by the explicit interior solutions that are found analytically in Appendix \ref{App:Interior}, which are derived under the assumption that the axio-dilaton varies only very slowly over the interior of the source. 
\end{itemize}

\begin{figure}[h]
\centering
\begin{subfigure}{.33\textwidth}
  \centering
  \includegraphics[width=.9\linewidth]{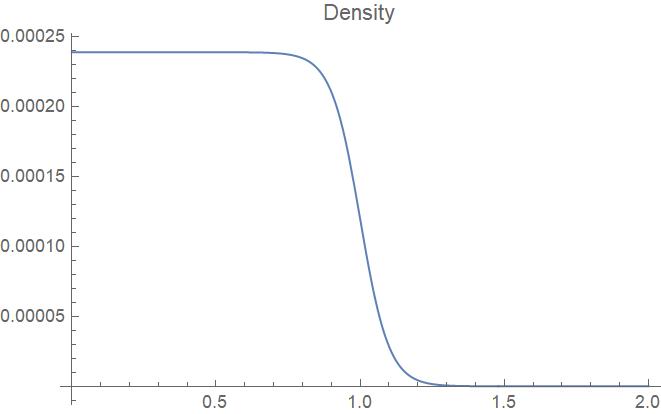}
  \caption{}
  \label{fig:sub1}
\end{subfigure}%
\begin{subfigure}{.33\textwidth}
  \centering
  \includegraphics[width=.9\linewidth]{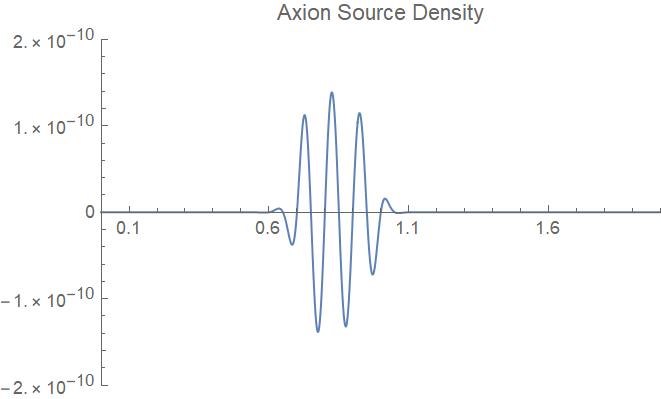}
  \caption{}
  \label{fig:sub2}
\end{subfigure}
\begin{subfigure}{.33\textwidth}
  \centering
  \includegraphics[width=.9\linewidth]{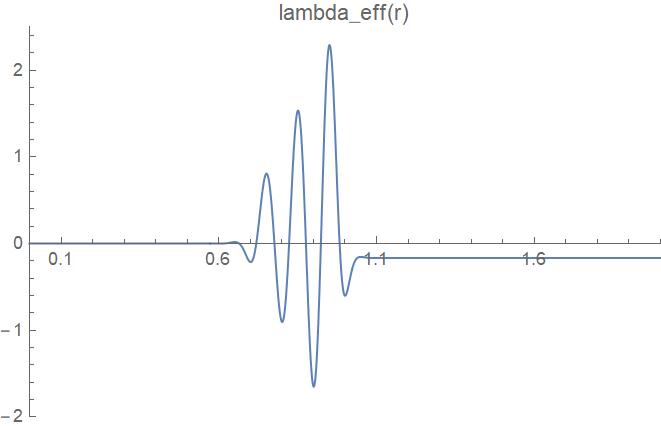}
  \caption{}
  \label{fig:sub9}
\end{subfigure}\\
\begin{subfigure}{.33\textwidth}
  \centering
  \includegraphics[width=.9\linewidth]{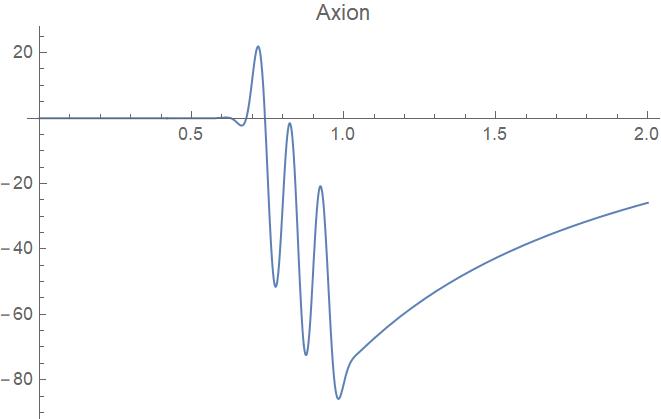}
  \caption{}
  \label{fig:sub3}
\end{subfigure}%
\begin{subfigure}{.33\textwidth}
  \centering
  \includegraphics[width=.9\linewidth]{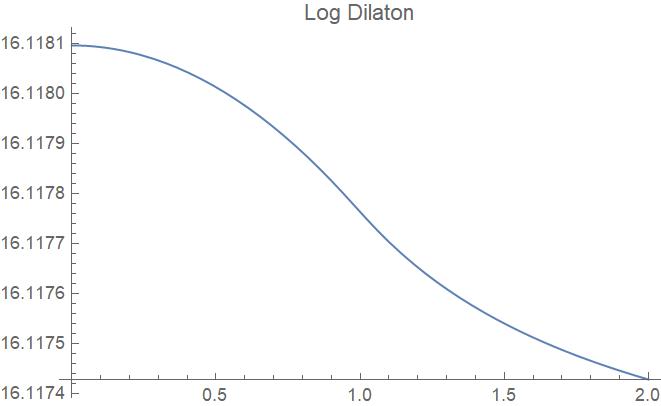}
  \caption{}
  \label{fig:sub4}
\end{subfigure}
\begin{subfigure}{.33\textwidth}
  \centering
  \includegraphics[width=.9\linewidth]{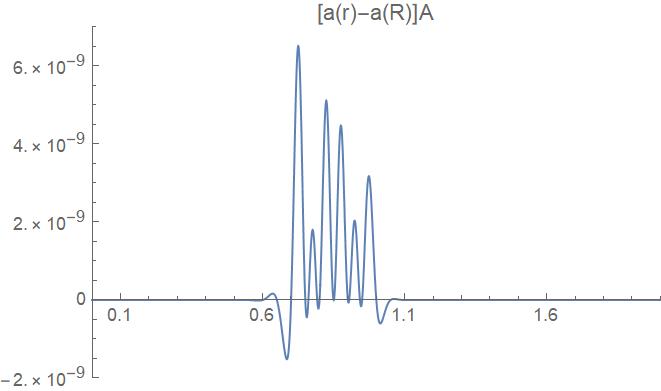}
  \caption{}
  \label{fig:sub5}
\end{subfigure}\\
\begin{subfigure}{.33\textwidth}
  \centering
  \includegraphics[width=.9\linewidth]{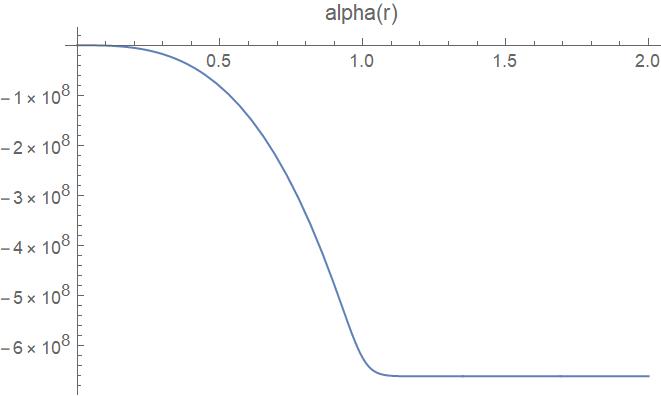}
  \caption{}
  \label{fig:sub6}
\end{subfigure}%
\begin{subfigure}{.33\textwidth}
  \centering
  \includegraphics[width=.9\linewidth]{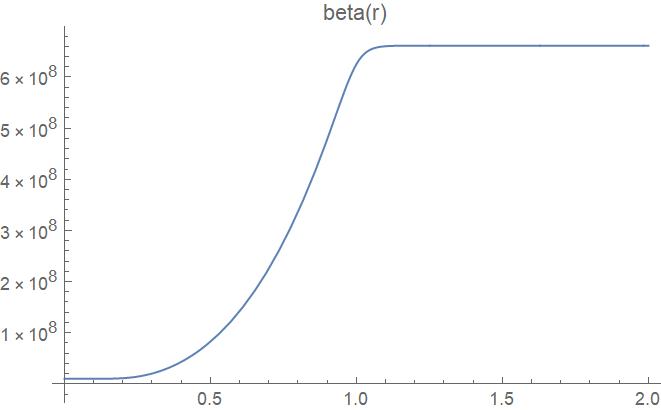}
  \caption{}
  \label{fig:sub7}
\end{subfigure}
\begin{subfigure}{.33\textwidth}
  \centering
  \includegraphics[width=.9\linewidth]{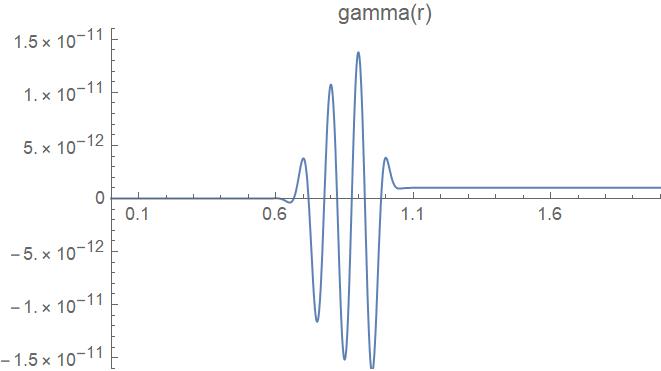}
  \caption{}
  \label{fig:sub8}
\end{subfigure}
\caption{Axio-dilaton properties as a function of $r/R$, showing the two input functions (a) $\rho(r)$ and (b) $\cA(r) = \varepsilon \rho$ together with the calculated outputs (c) $\lambda_{\rm eff}(r)$, (d) $\mfa(r)$, (e) $\log \tau(r)$, (f) $[\mfa(r) - \mfa(R)] \cA(r)$, (g) $\alpha(r)$, (h) $\beta(r)$ and (i) $\gamma(r)$. Here $\gamma(r)$ and $\alpha(r)$ are found by evaluating \pref{Aconsradint2} and \pref{Sconsradint2} at an arbitrary radius (rather than $r=R$) while $\beta^2(r) := \tau^2(r) + [\mfa(r) - \alpha(r)]^2$ and $\lambda_{\rm eff}(r) = -\beta(r) \gamma(r)/4GM$ with $M = 4\pi\int_0^R r^2 \rho(r) \, \exd r$. These must (and do) approach the constants $\alpha$, $\beta$, $\gamma$ and $\lambda_{\rm eff}$ exterior to the source. For this example $\lambda_{\rm eff} \simeq -0.168$, although other choices give positive $\lambda_{\rm eff}$.}
\label{Fig:Num}
\end{figure}

Suppression of the effective dilaton coupling therefore requires $\varepsilon$ to change sign within the source and requires the axion to vary quickly within the interior. We have searched this part of parameter space by evaluating the interior solutions numerically, seeking a functional form for $\varepsilon$ that allows $|\lambda_{\rm eff}|$ to be small, with particular interest in configurations where $[\mfa(r) - \mfa(R)] \varepsilon(r)$ can negative. Fig.~\ref{Fig:Num} shows a representative example of one such a numerical solution, chosen so that $\varepsilon$ changes sign multiple times near the source's surface. 

We have so far neither found an example with $|\lambda_{\rm eff}| < \frac16$ and -- although the numerical evidence suggests it -- have not yet been able to prove that this is impossible. 

\section{Conclusions}
\label{sec:Conclusions}

In this article we study the phenomenological aspects of an axion-dilaton system that appears generically in string and supergravity theories as real and imaginary parts of scalar components of gravitationally coupled chiral superfields or moduli. Recent studies suggest the potential importance of these fields  to be active for late-time cosmology and address important questions, such as the relation between the scales of dark energy, neutrino masses and electroweak symmetry breaking, if they happen to be light enough \cite{YogaDE}. In particular the presence of the light axion partner of the dilaton was suggested to provide a new mechanism to potentially  screen dilaton couplings to standard model matter -- axion homeopathy \cite{Homeopathy} --that could  help to evade the observational  problems of Brans-Dicke scalars. 

We examine both IR and UV aspects of these theories in turn. 

\subsection*{UV implications}

In the UV side we address the following issues:

\begin{itemize}
\item
{\it Axion UV coupling problem.} The large value of the dilaton field $\tau\simeq 10^{28}$ that provides the observed hierarchies of scale also apparently gives rise to a very low axion decay constant: $f\sim$ eV. This suggests strongly coupled axion dynamics at low enough energies to be dangerous. We explore the breakdown of EFT methods at eV scales by embedding into a concrete UV completion at these scales in terms of a higher dimensional supersymmetric theory with two very large extra dimensions. The supersymmetry of the gravity sector in this case arises because the extra-dimensional bulk is supersymmetric, and this is consistent with the Standard Model sector not being supersymmetric up to TeV scales if it is situated on a supersymmetry breaking brane (as is common in standard brane-world scenarios). Axions arise in this framework as KK modes for bulk antisymmetric tensor fields,  $B_{\ssM\ssN}$, and we verify that dimensional reduction reproduces the $SL(2,R)$-invariant kinetic terms whose presence predicts small $f$ (and so the need for UV completion).

\item
{\it Axion unitarization.} The UV completion allows us to identify the real coupling strength of each axionic mode, to compare with the naive EFT expectation. We find in all cases that individual axion states couple only with 4D Planck strength and so are not strongly coupled at any energies below TeV scales. We identify several different ways this happens, focussing on two in particular: `$S$-type' axions are dual to the universal 4-dimensional antisymmetric tensor $B_{\mu\nu}$ and  `$T$-type' axions come from components $B_{mn}$ in the two large directions of an asymmetric compactification. We find that the naive EFT coupling estimate for $T$-type axion-matter interaction gives $\cF \sim M_p$ rather than $\cF \sim f$ once carefully done. The naive estimate for couplings of the $S$-type axion in this case is correct -- $\cF \sim f$ -- but the UV theory instead produces its dual $b_{\mu\nu}$ which is only weakly coupled (again with Planck strength).
 
 \item{\it UV observational bounds.} Having a description at up to TeV energies allows a discussion of axion production and energy-loss bounds, and we find that these generally coincide with standard energy-loss constraints for higher-dimensional models. They need not pose a problem provided that the 6D gravity scale is above several tens of TeV.  This automatically also ensures that collider signals are too small to have yet been seen. 
 
 \item
{\it Microscopic origin of scales.} In the UV theory the large size of the dilaton field $\tau$ arises from the large size of the extra-dimensional volume in units of the 6D gravitational scale. Ref.~\cite{YogaDE} argues that the naive correspondence between them $\tau \simeq \cV^{2/3}$ cannot be valid because it would imply unrealistically small string and Kaluza-Klein scales. Ref.~\cite{YogaDE} proposed addressing this issue using stringy tools such as asymmetric compactifications, warped factors and superpotential tuning. We here provide a different mechanism in which additional factors of volume are contributed by 6D-dilaton and complex structure moduli, motivated by string-inspired examples. We leave for the future extending the theory about TeV scales through a proper string embedding including stabilization of all other moduli.

\item
{\it Yoga vs SLED Scenarios.} The UV completion we suggest is very similar to the scenario of supersymmetric large extra dimensions scenario (SLED) since in both cases Standard Model physics is localized on nonsupersymmetric branes within two large and comparatively supersymmetric extra dimensions. Because of this Yoga models inherit SLED properties at energies larger than the eV threshold and shares their nice UV properties (such as a decoupling of large on-brane UV scales from 4D spatial curvature). However at low energies the accidental low-energy symmetries of the Yoga scenario leads to smaller scalar potentials than were found in SLED models, suggesting that SLED scenarios might improve if moduli were stabilized to exploit the low-energy extended no-scale structure that Yoga models build in. 
\end{itemize}

\subsection*{IR implications}

In the IR we find

\begin{itemize}
\item
{\it New class of background solutions.}  We find a broad new class of solutions for the coupled axio-dilaton and Einstein field equations that apply outside a gravitating object. These solutions promote an arbitrary solution of a Klein-Gordon field coupled to gravity into a full solution of the Einstein-axiodilaton equations. We use these to extend the solutions found in \cite{Homeopathy} beyond spherical symmetry to include the general multipole expansion in the weakly gravitating regime. It is the underlying $SL(2,R)$ symmetry of the equations that allows such a general analytic result. We also explore solutions in the interior of the macroscopic object and match it to exterior solutions.
 \item
 {\it Breaking of shift symmetry.} We show that dilaton-matter couplings are not screened by axion-matter couplings if these do not break the axion shift symmetry within matter. If matter couplings do not break the axion shift symmetry then only higher multipole moments arise for the external axion field and we show why these fall off too quickly to contribute to the PPN parameters $\gamma_{\PPN}$ and $\beta_\PPN$ and so leave the standard Brans-Dicke result that is experimentally ruled out. 

\item{\it Numerical exploration of axio-dilaton response.} We numerically explore solutions when shift symmetry is broken by axion-matter couplings without also being broken in the vacuum. We find constraints the axion coupling that are necessary (but not sufficient) conditions for reducing apparent dilaton couplings to matter. We numerically explore axio-dilaton response to macroscopic gravitating sources for a wide class of axion-matter couplings, finding no examples that successfully suppress the effective dilaton-matter coupling\footnote{In \cite{Homeopathy}  interior solutions were left for future work. Here, we are including interior solutions and so far have not found the matching that effectively screens the Brans-Dicke field couplings to matter. We leave for future work a more complete study of screening mechanisms.}. 
\end{itemize}

Both very light axions and very light dilaton fields have been considered over the years for applications to astrophysics and late time cosmology. We find that considering both at the same time -- particularly with the $SL(2,R)$-invariant target-space interactions -- offers an even richer phenomenology that paradoxically evades some of the bounds each could individually suffer from separately. The fact that the required interactions also appear naturally in supersymmetric field and string theories only adds to their potential relevance. They bring many surprises: smaller vacuum energies and rich and unusual long-distance response. We leave further studies, addressing  screening mechanisms and including cosmological implications to a future publication \cite{philippe}

Although the suppression mechanism underlying Yoga models does allow smaller than usual vacuum energies, they so far seem to founder because the light scalars they imply seem to be ruled out by tests of gravity within the solar system. We believe the model-building issues needed to evade these tests are likely easier to solve than has been the cosmological constant problem itself, and although we have found a number of new ways in which axiodilaton-matter couplings can interestingly suppress how the dilaton couples to gravitating objects,\footnote{These include mixing of $S$, $T$ and $\Phi$ type axions and the interplay of their matter interactions with those of the Yoga model's relaxon field.} none yet have completely allowed solar system tests to be evaded. We find many more possibilities in the model-building of axiodilaton response to macroscopic objects than seem to be available as alternatives to understanding the cosmological constant problem. We hope Yoga models will prove to be a small step in the thousand-mile journey towards solving the puzzle of how the vacuum gravitates.

.

\section*{Acknowledgements}
We thank Mustafa Amin, Kevin Kelly, Justin Khoury, Azadeh Maleknejad, Francesco Muia, Andreas Schachner and Mark Wise for helpful conversations. CB's research was partially supported by funds from the Natural Sciences and Engineering Research Council (NSERC) of Canada. Research at the Perimeter Institute is supported in part by the Government of Canada through NSERC and by the Province of Ontario through MRI. The work of FQ has been partially supported by STFC consolidated grants ST/P000681/1, ST/T000694/1. 

\begin{appendix}

\section{Slowly varying interior solutions}
\label{App:Interior}

This appendix considers the simplest model for interior axion-matter couplings in order to illustrate in a solvable situation precisely how axion-matter couplings modify naive Brans-Dicke expectations and to identify more clearly how weak couplings approach the Brans-Dicke limit. Throughout it is important to remember that the sources $\rho(r)$ and $\cA(r)$ are also in general functions of the two scalar fields $\tau$ and $\mfa$, and so are in particular generically position dependent. The limit of an incompressible medium (with constant energy density) makes the most sense for weakly gravitating objects in the Jordan frame.

\subsection*{Matching}

As described in the main text, the general exterior solution \pref{semicircleAnsatz} is characterized by four integration constants ($\alpha$, $\beta$, $\gamma$ and $\delta$). Two of these ($\gamma$ and $\alpha$) are determined by the interior properties of the source through the matching conditions \pref{Aconsradint2} and \pref{Sconsradint2}, reproduced for convenience here:
\be \label{Aconsradint2app}
   \gamma =  R^2 \left( \frac{\partial_r \mfa}{\tau^2} \right)_{r=R} = - \frac{1}{3M_p^2}\int_0^R \exd r \; r^2 \cA(r)   \,,
\ee
and
\be \label{Sconsradint2app}
   \gamma \alpha = R^2 \left( \frac{\partial_r \tau}{\tau} + \frac{\mfa \, \partial_r \mfa}{\tau^2}\right)_{r=R} = - \frac{1}{3M_p^2} \int_0^R \exd r\; r^2 \Bigl[\rho(r) + \mfa(r) \,\cA(r) \Bigr] \,,
\ee
where $r = R$ is the radius of the source's surface. Using \pref{Aconsradint2app} in \pref{Sconsradint2app} allows it to be rewritten as
\be \label{Sconsradint2appv2}
   \gamma \Bigl[\mfa(R) -  \alpha \Bigr] = -R^2 \left( \frac{\partial_r \tau}{\tau}  \right)_{r=R} =  \frac{1}{3M_p^2} \int_0^R \exd r\; r^2 \Bigl[\rho(r) + [\mfa(r) - \mfa(R)] \,\cA(r) \Bigr] \,.
\ee
It turns out that if $\cA(r)$ has a definite sign then $\mfa(r < R) - \mfa(R)$ has the same sign as $\cA$ -- as can be seen from \pref{Aconsradmxy} below assuming the spherically symmetric initial condition $r^2\partial_r\mfa = 0$ at $r = 0$) -- and so eq.~\pref{Sconsradint2appv2} implies $\gamma[\mfa(R) - \alpha ] = \frac23 GM(1 + \Delta)$ with $\Delta \ge 0$.

The other two constants ($\beta$ and $\delta$) are determined by the values taken by the exterior fields, $\tau_\star = \tau(r_\star)$ and $\mfa_\star = \mfa(r_\star)$, at any fixed exterior reference radius $r_\star$ -- such as asymptotically at infinity ($r_\star = \infty$) or near the surface of the source ($r_\star = R$). Any such values determine the semicircle on which the exterior solution lies, so $\beta$ is determined by
\be
   \beta = \Bigl[ \tau_\star^2 + (\mfa_\star - \alpha)^2 \Bigr]^{1/2} \,,
\ee
while $\delta$ can be found from either of the following two conditions:
\be
   \tau_\star = \frac{\beta}{\cosh \delta} \quad \hbox{and} \quad \mfa_\star = \alpha + \beta \tanh \delta \,.
\ee
 
\subsection*{Interior field equations}

Consider first a spherically symmetric ansatz that breaks shift symmetry, possibly down to a discrete subgroup (as would be true in particular if it is the QCD anomaly that breaks the shift symmetry). An unbroken discrete symmetry implies physics is periodic under $\mfa \to \mfa + 2\pi$, with for instance $\rho(\mfa + 2\pi) = \rho(\mfa)$ and $\cA(\mfa + 2\pi) = \cA(\mfa)$. It can be useful to define $\varepsilon(\mfa,\tau) := \cA(\mfa,\tau)/\rho$, and in the case where $\mfa$ is a pseudoscalar $\varepsilon$ must be CP-violating and this implies $\varepsilon$ is even under $\mfa \to - \mfa$.

In this case for $r < R$ the spherically symmetric axion field equation \pref{AxDilEOM} is the differential version of \pref{Aconsradint2app}:
\be \label{Aconsradmxy}
    \partial_r \left( \frac{r^2 \partial_r \mfa}{\tau^2} \right)  = -   \frac{r^2\cA}{3M_p^2}  \,,
\ee
and the equation expressing conservation for $J_s^\mu$ in eq.~\pref{JEOMs} is similarly 
\be \label{Sconsradvxy}
    \partial_r \left[ r^2 \left( \frac{\partial_r\tau}{\tau} + \frac{\mfa \,\partial_r \mfa}{\tau^2} \right) \right] = - \frac{r^2}{3M_p^2} (\rho +\mfa \,\cA)   
\ee
and so
\be \label{Sconsradvxy2}
   \partial_r\left( \frac{r^2 \partial_r \tau}{\tau} \right) = -\partial_r\left( \frac{r^2 \mfa\, \partial_r \mfa}{\tau^2} \right)    - \frac{r^2}{3M_p^2} (\rho +\mfa \,\cA)   
   =  -  \frac{r^2\rho }{3M_p^2}  - \left( \frac{r \partial_r \mfa}{\tau} \right)^2    \,,
\ee
where the last equality uses \pref{Aconsradmxy}. 

We also require the metric equations and for weakly gravitating systems we can write the Einstein-frame metric as $g_{tt} \simeq - [1 + 2 \Phi(r)]$ where the Einstein equations imply $\Phi$ satisfies 
\be
  \nabla^2 \Phi \simeq 4\pi G \rho = \frac{\rho}{2M_p^2} \,,
\ee
and so when $\rho = M \, \delta^3(\bfr)$ we have the boundary condition $\bfn \cdot \nabla \Phi \simeq GM$ where $M := \int \exd^3\bfr \, \rho$, and so $\Phi \simeq -GM/r$ exterior to a spherically symmetric source. This shows for weak fields that the gravitational (ADM) mass agrees with the inertial mass in Einstein frame, with
\be 
   M := \int \exd^3x \; \rho = 4\pi \int_0^R \exd r \, r^2 \rho(r) \,.
\ee

$M$ is {\it not} the gravitational mass, $M_g = \widetilde M$, as measured using the motion of orbiting test bodies, since this is instead given by the $1/r$ term in the Jordan-frame metric component $\tilde g_{tt} \simeq -[1 - 2G\widetilde M/r+\cdots]$. Given the large-$r$ expansion of $\tau$ given in \pref{LargerAsympt1},  
\be  \label{LargerAsympttauapp}
   \tau  =  \frac{\beta}{\cosh X} = \tau_\infty \left[ 1 - \frac{\beta \gamma}{r} \, \tanh \delta + \cdots \right]   \,,
\ee
we see that
\be
   -\tilde g_{tt} = -A^2 g_{tt} = -\frac{\tau_\infty}{\tau} \, g_{tt} = \left(1 + \frac{\beta\gamma}{r} \, \tanh \delta + \cdots \right) \left(1 - \frac{2GM}{r} + \cdots \right)
\ee 
and so
\be
  \widetilde M \simeq M \left(1 - \frac{\beta\gamma}{2GM} \tanh\delta \right) = M \left(1 + 2 \lambda_{\rm eff} \tanh\delta \right) 
\ee
which uses the definition \pref{BDeffDef}: $\lambda_{\rm eff} = - {\beta \gamma}/({4GM})$. 

\subsection*{Frame-dependence of the equation of state}

Before comiting to an equation of state for the interior it is useful to distinguish how the sources are related to one another in Jordan and Einstein frames. In particular, incompressibility is most naturally postulated in Jordan frame if the physics responsible comes from ordinary particles.

Using the definitions $T^{\mu\nu} = (2/\sqrt{-g}) (\delta S_m /\delta g_{\mu\nu})$ and $\widetilde T^{\mu\nu} = (2/\sqrt{- \tilde g}) (\delta S_m /\delta \tilde g_{\mu\nu})$ for the Einstein- and Jordan-frame stress tensors, together with $\tilde g_{\mu\nu} = A^2 g_{\mu\nu}$ implies $\sqrt{-\tilde g} = A^4 \sqrt{-g}$ and so
\be
 \widetilde T^{\mu\nu} = A^{-6} T^{\mu\nu} \,, \quad
  {\widetilde T_\mu}{}^{\nu} = A^{-4} { T_\mu}{}^{\nu} \,, \quad
 \widetilde T_{\mu\nu} = A^{-2} T_{\mu\nu} \,, 
\ee
where indices are raised and lowered using the corresponding metric. Similarly the Jordan- and Einstein-frame fluid 4-velocities satisfy $\tilde g_{\mu\nu} \widetilde U^\mu \widetilde U^\nu = g_{\mu\nu} U^\mu U^\nu = -1$ and so $\widetilde U^\mu =A^{-1} U^\mu$. Finally, defining energy density and pressure in both frames using $T^{\mu\nu} = (p+\rho) U^\mu U^\nu + p \, g^{\mu\nu}$ and $\widetilde T^{\mu\nu} = (\tilde p+\tilde \rho) \widetilde U^\mu \widetilde U^\nu + \tilde p \, \tilde g^{\mu\nu}$ we see the Jordan- and Einstein-frame pressure and energy density are related by\footnote{Notice that scale factors are related by $\tilde a = A a$ and so JF and EF particle densities satisfy $\tilde n \tilde a^3 = n a^3$ and so $\tilde n = A^{-3} n$. Then the equation of state for dust becomes $\tilde \rho = \tilde m \tilde n$ in JF and $\rho = m n$ in EF and the consistency of these with $\tilde\rho = A^{-4}\rho$ implies $m = \tilde m A$, precisely as required if $\tilde m$ were independent of $\tau$ and $m \propto \tau^{-1/2} \propto A$.} $\tilde p = A^{-4} p$ and $\tilde \rho = A^{-4} \rho$. In particular, equation of state parameters like $w = \tilde p/\tilde \rho = p/\rho$ are the same in both frames and incompressible fluids in Jordan frame have constant $\tilde \rho$ and so 
\be
  \rho = \tilde \rho A^4 = \tilde \rho \left( \frac{\tau_\infty}{\tau} \right)^2 \propto \tau^{-2} \,.
\ee

\subsection*{Interior solutions with broken shift symmetry}

Eqs.~\pref{Aconsradmxy} and \pref{Sconsradvxy2} can be solved for the internal profiles $\tau_{\rm int}(r)$ and $\mfa_{\rm int}(r)$ by changing variables to $u$ as defined by 
\be \label{udefeq}
    u^2 :=  \frac{2GMr^2}{R^3} \quad\hbox{for which }\quad  0 < r < R \quad\hbox{implies} \quad 0 < u^2 < \frac{2GM}{R} \ll 1 \,,
\ee
and writing $\rho(r) = [3M/(4\pi R^3)] \hat \rho(u)$ where $\hat \rho$ is dimensionless and is normalized to satisfy $\int \exd^3x \, \hat \rho = \frac43 \pi R^3$. In terms of this, and writing $\cA = \varepsilon\, \rho$, eqs.~\pref{Aconsradmxy} and \pref{Sconsradvxy2} become
\be \label{Aconsradmxyu}
    \partial_u \left( \frac{u^2 \partial_u \mfa}{\tau^2} \right)  = -  u^2\varepsilon \hat \rho \,,
\ee
and
\be \label{Sconsradvxy2u}
   \partial_u\left( \frac{u^2 \partial_u \tau}{\tau} \right) +  \left( \frac{u\, \partial_u \mfa}{\tau} \right)^2    
   =  -  u^2\hat\rho  \,.
\ee

In principle $\hat\rho$ is fixed by solving the matter field equations expressing hydrostatic equilibrium (which, for gravitationally bound systems, can involve a feedback on the local gravitational potential $\Phi$). We take the simplest situation: where the Jordan-frame density is constant (incompressible material) -- and so $\hat\rho = (\tau_{\rm ref}/\tau)^2$, with $\int \exd^3x \, \tau^{-2} =: \frac43 \pi R^3\tau_{\rm ref}^{-2}$ -- as also is the proportionality function $\varepsilon$. Then eqs.~\pref{Aconsradmxyu} and \pref{Sconsradvxy2u} become
\be \label{Aconsradmxyuincomp}
    \partial_u \Bigl( u^2 \partial_u \mfa \Bigr) - \frac{2  u^2 \partial_u \tau\, \partial_u \mfa}{\tau}   = -  u^2\varepsilon \tau_{\rm ref}^2 \,,
\ee
and
\be \label{Sconsradvxy2uincomp}
   \partial_u\left( \frac{u^2 \partial_u \tau}{\tau} \right) +  \left( \frac{u\, \partial_u \mfa}{\tau} \right)^2    
   =  -  \left( \frac{u \tau_{\rm ref}}{\tau} \right)^2  \,.
\ee

Because $u$ is always small within a weakly gravitating source slowly varying solutions can be found as a series in powers of $u$, which starts at order $u^2$ because smoothness at the origin requires the first radial derivative to vanish at $r = u = 0$. So for $0 < r < R$ we seek solutions of the form:
\be
   \tau_{\rm int} = \tau_0 \Bigl( 1 + c_2 u^2 + \cdots \Bigr) \quad \hbox{and} \quad
    \mfa_{\rm int} = \mfa_0 + a_2 u^2 + \cdots   \,,
\ee
in which case eqs.~\pref{Aconsradmxyuincomp} and \pref{Sconsradvxy2uincomp} imply $c_2 = - \frac16(\tau_{\rm ref}/\tau_0)^2$ and $a_2 = - \frac16 \tau_{\rm ref}^2 \varepsilon(\mfa_0,\tau_0)$, so
\be
    \mfa_{\rm int} = \mfa_0 - \frac{\varepsilon \tau_{\rm ref}^2GM}{3R} \left( \frac{r}{R} \right)^2  + \cdots \quad
    \hbox{and} \quad
    \tau_{\rm int} = \tau_0 \left[ 1 - \frac{GM \tau_{\rm ref}^2}{3R\tau_0^2}  \left( \frac{r}{R} \right)^2  + \cdots \right] \,,
\ee
This allows us to compute $\tau_{\rm ref}$ as a function of $\tau_0$, $M$ and $R$ using
\be
 \frac{1}{\tau_{\rm ref}^2} = \frac{3}{4\pi R^3} \int \exd^3x \; \frac{1}{\tau^2_{\rm int}} = \frac{3}{R^3} \int_0^R \exd r\; \frac{r^2}{\tau^2_{\rm int}(r)} \simeq \frac{1}{\tau_0^2} \left( 1 + \frac{2GM \tau_{\rm ref}^2}{5 R \tau_0^2} \right)  \,,
\ee
and so $\tau_{\rm ref} \simeq \tau_0$ up to $GM/R$ corrections.
 
Using these to match to the exterior solutions $\tau = \beta/\cosh X$ and $\mfa = \alpha + \tanh X$, with $X = (\beta\gamma/r) + \delta$, the constants $\gamma$ and $\alpha$ become
\be \label{gammaperiodic}
   \gamma = R^2 \left( \frac{\partial_r\mfa_{\rm int}}{\tau_{\rm int}^2} \right)_{r=R} \simeq  - \frac{2 GM}{3}\, \varepsilon(\mfa_0)  \,,
\ee
and 
\be \label{Sconsradx2y}
  \gamma \alpha  =  R^2 \left( \frac{\partial_r\tau_{\rm int}}{\tau_{\rm int}} + \frac{\mfa_{\rm int} \,\partial_r \mfa_{\rm int}}{\tau^2_{\rm int}}\right)_{r=R} \simeq  - \frac{2 GM}{3}  -  \frac{2GM}{3} \, \varepsilon (\mfa_0) \; \mfa(R) \,.
\ee
Eliminating $\gamma$ from these last two gives
\be \label{logtauprimeBCmx2y}
 \alpha \simeq\mfa(R) + \frac{1}{\varepsilon(\mfa_0)}   \,.
\ee
Notice that these solutions ensure $\mfa(r) - \alpha$ is a periodic function of $\mfa_0$ everywhere outside the source. 

$\beta$ and $\delta$ are determined using continuity with the exterior solutions, leading to
\be
   \tau(R) = \tau_0 \left( 1 - \frac{GM}{3R} + \cdots \right) =  \beta\, \hbox{sech} \left( \frac{\beta \gamma}{R} + \delta \right)\,,
\ee
and
\be
   \mfa(R) \simeq \mfa_0 - \frac{\varepsilon(\mfa_0) GM}{3R}+ \cdots = \alpha  + \beta \, \tanh \left( \frac{\beta\gamma}{R} + \delta \right)   \,.
\ee
These imply that the radius of the semicircle defined by the fields at $r = R$ is (using \pref{logtauprimeBCmx2y})
\be\label{betaperiodic}
   \beta = \Bigl[ \tau^2(R) + [ \mfa(R) - \alpha ]^2 \Bigr]^{1/2} \simeq \left[\tau_0^2 \left(1 - \frac{2GM}{3R} \right) +  \frac{1}{\varepsilon^2(\mfa_0)}  \;\right]^{1/2} \,.
\ee
Together with \pref{gammaperiodic} eq.~\pref{betaperiodic} gives the effective Brans-Dicke parameter, through the relation
\be
   \lambda_{\rm eff} = - \frac{\beta\gamma}{4GM} = \frac{\beta}{6} \,\varepsilon(\mfa_0) \simeq \frac16   \left[1 + \tau_0^2\varepsilon^2(\mfa_0) \left(1 - \frac{2GM}{3R} \right) \right]^{1/2}\,.
\ee
This approaches the naive Brans-Dicke coupling $\lambda_{\rm eff} \to \frac1{6}$ (for all $\mfa_0$) if $\varepsilon \to 0$ with all other fields fixed basically because \pref{betaperiodic} implies $\beta \to 1/\varepsilon$ in this limit. Notice also that $\lambda_{\rm eff}$ is strictly larger than its Brans-Dicke value and only starts to differ appreciably from it once $\varepsilon(\mfa) \gsim 1/\tau_0$.

The value of $\delta$ is then found using, for instance
\be
   \mfa(R) - \alpha = -\frac{1}{\varepsilon(\mfa_0)} = \beta \, \tanh\left( \frac{\beta\gamma}{R} + \delta \right) = -\frac{6 \lambda_{\rm eff}}{\varepsilon(\mfa_0)} \, \tanh\left( \frac{4 \lambda_{\rm eff} GM}{R} - \delta \right) \,,
\ee
and so $\tanh\delta \simeq 1/(6 \lambda_{\rm eff})$ (which is smaller than unity because $\lambda_{\rm eff} \ge \frac16$). Given these integration constants the fields at infinity are given in terms of $\tau_0$ and $\mfa_0$ (or vice versa) by
\be
  \frac{1}{\tau_\infty} =\frac{\cosh\delta}{\beta} \quad \hbox{and} \quad
  \mfa_\infty = \alpha - \beta \tanh \delta  \,.
\ee

\end{appendix}

\end{document}